\def\Roman#1{\uppercase\expandafter{\romannumeral#1}}
\documentclass[a4paper,12pt]{article}
\usepackage{amssymb,amsfonts}
\usepackage[mathscr]{eucal}

\title{Path integral measure factorization  in  path integrals for diffusion of  Yang--Mills fields}
\author{S. N. Storchak\\ 
\small{Institute for High Energy Physics, Protvino, Moscow Region,142284,Russia}}

\begin{document}


\maketitle

\begin{abstract}
Factorization of the (formal) path integral measure in 
a Wiener path integrals for Yang--Mills diffusion is studied.

Using the nonlinear filtering stochastic differential 
equation, we perform the transformation of the path integral 
defined on a total space of the Yang--Mills principal fiber bundle
and come to
the reduced path integral on a Coulomb gauge surface.

Integral relation between the path integral representing the ``quantum'' evolution given on the original manifold of Yang--Mills fields and the path integral on the reduced manifold defined by the Coulomb gauge is obtained.

\end{abstract}

\section{Introduction}

The gauge field theories 
belong to a class of  infinite dimensional dynamical systems
with a symmetry.
One of the main problem in theoretical and mathematical
physics is the quantization of such dynamical systems.
A crucial question is of how   the extra degrees of freedom 
should be treated in quantization of the gauge theories.

Every dynamical system with a symmetry give rises
to a system with a lower degrees of 
freedom. A new system (a reduced dynamical system) may be
completely described  via  invariant variables.

It can be assumed that  similarly to the classical case  
a quantum behaviour
of two systems (the original system and the reduced one)
are also related to each other. 
This assumption is verified, for example,
in path integral quantization of finite dimensional dynamical systems 
with a symmetry where we have  
an integral relation between the corresponding path integrals
for the original and the reduced dynamical systems.

In order to get the path integral for the reduced system  in a gauge field theory, 
we make use of the Faddeev -- Popov 
method \cite{Faddeev}.
In this method  we consider the quantum
evolution on a gauge surface. This evolution is equivalent to the quantum evolution of the
dynamical system given on the gauge orbit space.

At present, the Faddeev -- Popov method of the path integral
quantization of the gauge theories is 
 the most  effective method for  studying 
the reduced quantum evolution in the  perturbation theory.
In this quantization method, the special transformation of 
the path integrals cancels the redundent degrees of freedom 
that are related to the gauge symmetry.

Another method which can be used for description of the quantum
evolution of the reduced system was
 proposed by Rossi and Testa in \cite{Rossi}. 
Almost the same approach 
to the quantization of the Yang--Mills field was given in the paper of  
  Teitelboim  \cite{Teitelboim}.
In these methods, the quantum evolution on the  orbit space of a  group action was presented\footnote{This evolution is locally equivalent to the evolution  on a
gauge surface.} 
by the integral over the gauge group with 
the original gauge field propagator as the integrand.

Exploring the relationship between
the Faddeev--Popov quantization  and that one given 
by Rossi and Testa, we carried out  the model investigation of
a path integral reduction problem in a finite dimensional case
 \cite{Storchak_1,Storchak_2}.
 We have considered the dynamical system 
describing the motion of a scalar particle on a smooth
compact manifold with a given Lie group action.
This dynamical system may be regarded as a finite dimensional model
for a dynamical system with a gauge symmetry.
A standard quantization of this model was presented in \cite{Kunstatter,Tanimura}. 

In our papers, we have considered the diffusion of a scalar particle
on a smooth compact manifold
and have established that a path integral measure generated 
by the stochastic process  does not invariant under reduction.
The path integral transformation (from
the original path integral to the path integral which describes the evolution of  the reduced system)
gave rise  to   additional terms (the transformation
Jacobian) in the differential generator 
(the reduced  "quantum Hamiltonian'') 
of the semigroup related with the reduced stochastic process. 

This result was  obtained by factorization  of a path integral
measure. An initial path integral measure was decomposed 
into two    measures. The first measure was generated by the stochastic
process given on the orbit of the group action 
and  the second measure   was constructed by the
stochastic process defined on the orbit space.

The path integral measure transformation was performed 
by using the
nonlinear filtering stochastic differential equation from the
stochastic process theory.
Due to the  symmetry of  our model, 
the nonlinear filtering equation was in fact the linear 
equation. This allowed us to present its solution  via
 the multiplicative stochastic integral.

In this paper we  extend our method of the local measure
factorization \cite{Storchak_2} to the path integrals of the
Euclidean Yang--Mills field theory.
That is, considering the Schr\"odinger approach 
to the quantization of
the field theories, we  shall study the transformation of the path
integral which represents the evolution of an initial 
function given on a space of  Yang--Mills connections.
As in \cite{Rossi},   the noncanonical variable $A_0$ will be excluded
from  evolution by the gauge condition $A_0=0$.
The residual gauge degrees of freedom are related to the time independent  gauge transformations. 
We  shall fix this rest symmetry by  using   the Coulomb gauge.  

It is known that owing to an  ambiguity of the Coulomb gauge, one can not uniquely determine 
 the coordinates of a point on a total space of the principal fiber bundle.
Nevertheless we  assume in the paper that in  restricted 
domain, to which our evolution belongs, it is possible to  use this gauge for
fixing the residual degrees of  gauge  freedom.
 Because of the local character of our evolution   
we  shall not consider in the paper  the effects coming from
the nontrivial topology of the reduced manifold.

 In a finite dimensional case,
a free, proper and  isometric action of a Lie group on a smooth
compact manifold leads to  a principal fiber bundle 
picture in which an original manifold {\cal P} can be viewed
as a total space of this principal fiber bundle.
In a gauge theory, the original manifold  is an infinite dimensional
space of the Yang--Mills connections. In this case, in order
to meet the requirements of the slice theorem by which one 
gets the principal fiber bundle structure on a manifold,
one has to impose some
additional restrictions both on  gauge connections and on the gauge
transformation group.
These questions have been  studied in detail in \cite{Singer_Scrip,
 Narasimhan,Parker,Soloviev}. 
The results of these investigations enable  us by two possibilities.

According to the first possibility, one must   use the pointed gauge 
group with the elements from the corresponding Sobolev class of functions.
Elements of this group  are restricted to satisfy the  
 condition $g(x_0)=e$  at some fixed point $x_0$ of the manifold $M$. 
 (Here $e$ is an identity element of a gauge group.) 

We shall follow in the paper the second possibility by which the
points of a manifold $\cal P$  are chosen to be the irreducible 
connections in the principal fiber bundle $P(M,G)$ in  Sobolev class $H_{k}$, $k>3$.
Also, the quotient group of
the gauge transformation group by its center\footnote{The center of the gauge group is a stabilizer at each irreducible connections.} is used as the transformation group 
acting on this manifold. We shall denote this quotient group by $\cal G$, with $g\in \cal G$ of $H_{k+1}$. 
Then, by the slice theorem it can be proved   that
the orbit space  $\cal M=\cal P /\cal G$  is a Hilbert manifold.

Further restrictions that one has to impose onto the class of allowed gauge fields 
are related with  path integrations. 
 In a finite dimensional case, we have used  the path integrals in which measures
 were generated by the solutions of the stochastic differential equations. 
The main contribution to the study 
of the stochastic processes for Yang--Mills fields
has been done by Asorey and Mitter in \cite{Asorey}. 
In their papers, it was given a rigorous definition of the regularized stochastic process 
on a Hilbert manifold of the Yang--Mills orbits.
They proved   the   It\^o formula for the stochastic differentiation
of the Yang--Mills  stochastic processes.\footnote{This  formula for the processes in Banach and Hilbert spaces was also proved in \cite{Dalecky}.}
Besides, they  have constructed the regularized semigroup  and its infinitesimal generator for the quantum evolution 
on the orbit space. 

These results were obtained by  compactifying the three--dimensional space  with a volume cut--off and  introducing the ultraviolet regularization. 
An original plane  metric given on the space of gauge connections was modified by inserting of
the extra factors $({\Roman 1}+{\triangle}_A/{\Lambda}^2)^k$,  where ${\triangle}_A=d_A\,d^{\ast}_A+
d^{\ast}_A\,d_A$ is the Laplace operator and  ${\Lambda}$ is  a   cut--off parameter. 
Such a replacement of the original (weak)  metric 
allowed them 
to obtain a family of
gauge  invariant (strong) Riemannian metrics given on 
the space of the gauge connections $\cal P$.
Then,  using the regularized metric, they have constructed the diffusion  processes on the tangent 
space over the original manifold of the Yang--Mills connections and  on  the gauge orbit manifold $\cal M$. 

In \cite{Asorey}, the global stochastic processes on a manifold have been  defined from  local processes
by a standard method based on the  parallel displacement in the fiber bundle.
Notice that there is  
another approach to the global definition of the stochastic process 
on a  manifold. It  was proposed by Belopolskaya and Daletskii in \cite{Dalecky}. 
It is their method was used in our papers concerning the path integral reduction in the finite dimensional case. 
In this method  the local stochastic processes, defined on charts of the tangent bundle, 
are mapped onto the manifold with the help of the exponential mapping. 
It follows that  
one can  study the global stochastic evolution by means of the  local evolutions given on charts of the manifold.

The aim of our present investigation is to establish the relation between the path
integral which describes the quantum evolution given on the original
manifold of the Yang--Mills connections  and 
the path integral representing  the quantum evolution on the gauge orbit space. 
The investigation will be based on 
the approach to the path integral measure factorization developed earlier in \cite{Storchak_1,Storchak_2}.

We shall consider the 
case when the Riemannian metric of the original manifold (the manifold of gauge connections) is plane.
In this case in order to have a properly defined Gaussian measure  and the Wiener process on a tangent space to the original manifold
one has to describe an evolution  with the help of the rigged space: ${\cal H}_{+} \subset {\cal H} \subset {\cal H}_{-}$. 
A choice of the corresponding Sobolev classes of functions for 
these Hilbert spaces (the choice of a definite ${\cal H}_{-}$) is determined 
by the analytical restrictions that are necessary for the definition of  diffusion processes (and the path integrals) related with 
the original evolution given in the  principal fiber bundle.
The diffusion processes in such spaces were studied in \cite{Dalecky}.

We note that the inductive limit of the chain of continuously and densely embedded Sobolev--Hilbert spaces of distributions (the  
Hilbert spaces ${\cal H}_{-k}$) formed by 
completion of the Schwartz space of functions ${\mathscr S}$ with respect to the  appropriate  norms is 
the space ${\mathscr S}^{'}(R^3)={\bigcup}_{k=1}^{\infty}  {\cal H}_{-k}(R^3)$\footnote{${\mathscr S}^{'}$ is the Schwartz space of tempered distribution.} 
of the Schr\"odinger representation of the quantum field theory.  
The quantum operators of this representation  act in the space 
$L^2({\mathscr S}^{'},d\mu)$. In case of a free quantum field theory, the measure $\mu$ is a Gaussian measure. 

In path integral approach to the quantization of the field theories, 
one of the main  problems is a correct treatment of the interaction potential in the Feynman--Kac formula.
Before using this formula in the path integral describing the evolution in the space of functions given on ${\mathscr S}^{'}$,
one has to regularize the potential term of the Hamiltonian.  We  remark that  the consistent solution of the quantum evolution
 problem in this approach was only done for the  limited number of the simple quantum field models.

On the other hand, the study of the
quantum evolution with  the regularized form of the Hamiltonian (as e.g. in  \cite{Asorey})
results in   the renormalization
problem in the obtained  final expression. 
 We note that general solution of this problem in path integration is known at present only in the scope of the perturbation theory. 

By this reason in the paper we shall
study   the particular case of 
the evolution which is given on the Hilbert manifold 
of the gauge fields equipped 
with the plane (unregularized) metric.
 It means, in fact,  that 
we shall deal with the factorization of the formal path integral measure.

\section{Backward Kolmogorov equation}
The Feynman propagation kernel 
\[
G(A_b,t_b;A_a,t_a)=<A_b|{\rm e}^{-iH(t_b-t_a)}|A_a>
\]
in $A_0=0$ gauge can be written
symbolically as \cite{Rossi}:
\[
G(A_b,t_b;A_a,t_a)=\int\limits_{\scriptstyle A(t_a,{\mathbf x})
=A_a({\mathbf x})\atop \scriptstyle A(t_b,{\mathbf x})=
A_b({\mathbf x})}
D A(t,{\mathbf x})\;{\rm e}^{iS(A_0=0)}\,,
\]
where the action $S$ is 
\[
S=\int ^{t_b}_{t_a}dt\int d^3x \;L,
\]
with $L=\frac{1}{2g_0^2}\sum Tr(F_{\mu\,\nu}F^{\mu\,\nu})$,
$F_{\mu\,\nu}=\partial_{\mu}A_{\nu}-\partial_{\nu}A_{\mu}+
[A_{\mu},A_{\nu}]$.

We shall investigate  not a true quantum evolution 
given by  above Feynman path integral but the evolution   
generated by the corresponding 
diffusion process on the Riemannian manifold of Yang--Mills connections.
The transition probability of this process can be defined 
by the solution of the corresponding backward Kolmogorov equation. 
The Green function of the backward Kolmogorov equation with
the self--adjoint operator  also satisfies  the forward equation 
which, in turn, can be transformed   
into the Schr\"odinger equation for the
Feynman propagation kernel by changing the parameter $\kappa $ of the equation for $i$. 

Notice that  the transition from the Wiener path integrals representing the solution of the forward Kolmogorov equation to the Feynmann path integrals 
is an independent problem in path integration. 

Our backward Kolmogorov equation has the following formal form:
\begin{equation}
\left\{
\begin{array}{l}
\displaystyle
\left(
\frac \partial {\partial t_a}
+\frac 12\mu ^2\kappa \,\triangle
_{\cal P}[A_a]+\frac
1{\mu ^2\kappa }\,
V[A_a]\,\right){\psi}_{t_b} (A_a({\mathbf x}),t_a)=0\\
{\psi}_{t_b} (A_b({\mathbf x}),t_b)=\phi _0(A_b({\mathbf x})),
\qquad\qquad\qquad\qquad\qquad (t_{b}>t_{a})\,.
\end{array}\right.
\label{1}
\end{equation}
In eq.(\ref{1}), $A_a({\mathbf x})\equiv(A_a)^{\alpha}_i({\mathbf x})$,
$\mu ^2= \hbar g_0^2$, $\phi _0(A)$ is a given initial function 
of the gauge connection, $\kappa $ is a real positive parameter.
$\triangle _{\cal P}[A]$ is the Laplace operator  given on the original plane Riemannian manifold $\cal P$ of
  gauge connections:
\[
\triangle _{\cal P}[A]=
G^{({\alpha}, i,x)\;({\beta},j,x')}
\frac{{\delta}^2}{\delta A^{(\alpha ,i,x)}\;
\delta A^{(\beta ,j,x')}}=
\int d^3x \, k^{\alpha \beta}\delta_{ij}
\frac{{\delta}^2}{\delta A^{\alpha}_i({\mathbf x})\;
\delta A^{\beta}_j({\mathbf x})}\,,
\]
where $G^{({\alpha}, i,x)\;({\beta},j,x')}={\delta}^{\alpha\,\beta}\,
{\delta}^{i\,j}\,{\delta}^3({\mathbf x}-{\mathbf x}')$.
(Here and what is follows we assume  summation over 
equal  discrete indices and  integration 
in case of equal continuous indices.)
 An invariant potential term $V[A]$ of the Yang--Mills 
Hamiltonian is
\[
V[A]=\int d^3x \,\frac{1}{2}\,
k_{\alpha \beta}\,
F^{\alpha}_{ij}({\mathbf x})\,F^{\beta \;ij}({\mathbf x})\,,
\]
$k_{\alpha \beta}=c^{\tau}_{\mu \alpha}c^{\mu}_{\tau \beta}$ is the Cartan--Killing metric for a group G.

We see that owing to the singular behavior of  $G^{({\alpha}, i,x)\;({\beta},j,y)}$, 
there are two functional derivatives given at the same point
in the Laplace operator of eq.(\ref{1}).
It  may lead to  divergency and  the
expression, as it stands, is not correctly defined.

There are   various methods to get over this difficulty.
It can be done, for example,
by  modifying the original plane 
metric $G_{({\alpha}, i,x)\;({\beta},j,y)}$.
In \cite{Asorey}, 
an extra  
 convergent factor was introduced into the original weak metric.
Thereby the Hilbert manifold was equipped with a strong Riemannian metric   and it became possible to determine 
the regularized Laplace operator which acts in the space of twice differentiable and bounded functions given on this manifold.

When one  describes the evolution in   rigged Hilbert spaces 
the domain of the properly defined Laplace operator  is 
the space of functions given on the Hilbert space  of distributions 
${\cal H}_{-}$.\footnote{In case of the Schr\"odinger representation of the quantum field theory  
this operator acts in the space of functions on ${\mathscr S}^{'}$. }
 The action of the Laplace operator in 
this space
can be defined \cite{Dalecky1,Berezansky}  as  trace (in ${\cal H}$) of 
the second derivative of these functions taken  in  ${\cal H}_{-}$
and then restricted to  ${\cal H}$. The kernel of the corresponding differential operator  is usually written (in a  symbolical form) by means of the variational derivatives.
It is also possible to determine the Laplace operator 
by its action on cylindrical functions given on ${\cal H}_{-}$
(see e.g. \cite{Albeverio} or  \cite{Semmler}).

 We note that    if one chooses  a (weak) scalar products (as a basic scalar products)
 in the Hilbert spaces, that are the tangent spaces to the original manifold, then 
one  needs to come  to the rigged Hilbert spaces.  
The rigged Hilbert space given in the tangent space 
can be obtained by making use of the Hilbert space  construction 
from \cite{Asorey}.
The model  space of 
the  Hilbert manifold with the modified Riemannian metric obtained 
there can be taken as the Hilbert space ${\cal H}_{+}$
in the triple of the rigged spaces. This space and its adjoint space  ${\cal H}_{-}$
(with respect to ${\cal H}$)  supplies the tangent spaces to
the original manifold $\cal P$ with 
the rigged structure 
${\cal H}_{+} \subset {\cal H} \subset {\cal H}_{-}$.

The obtained picture should be  further generalized since    
in quantum field theory the gauge fields belong to the space of 
distributions.

Therefore, one has to deal with a manifold  which is 
a Hilbert  manifold modeled on the Sobolev--Hilbert space of distributions 
${\cal H}_{-}$.
An inductive limit of a chain of such manifolds could be regarded as ``manifold of Schwartz distributions''. This object is not yet elaborated enough to be applied in our case, but we note that it is possible to determine 
(see ref.\cite{Paycha})
the ``dual manifold`` which is the projective limit of the corresponding Hilbert manifolds.

According to \cite{Dalecky} the solution of 
the backward Kolmogorov equation 
for the diffusion on the Hilbert Riemannian manifold ${\cal P}$
(provided that the coefficients  of the equation are  appropriately chosen) can be presented 
as a limit (under subdivision of the time interval) of the
superposition of the local semigroups: 
\begin{equation}
\psi _{t_b}(A_a({\mathbf x}),t_a)={\hat U}(t_b,t_a)\phi _0(p_a)=
{\lim}_q {\tilde U}_{\eta}(t_a,t_1)\cdot\ldots\cdot
{\tilde U}_{\eta}(t_{n-1},t_b)
\phi _0(A_a({\mathbf x})).
\label{eq4}
\end{equation}
The local evolution semigroups ${\tilde U}_{\eta}$ are determined by the equations:
\begin{equation}
 {\tilde U}_{\eta}(s,t)\phi (A)={\rm E}_{s,A} \phi ({\eta}(t))
\,\,\,
s\leq t,\,\,\,\eta (s,{\mathbf x})=A({\mathbf x})\,,
\label{locsemi}
\end{equation}
where the expectation value of the functions $\phi $ 
is taken over the stochastic process which is a local representative of the global stochastic process ${\eta}_t $ 
obtained by means of the exponential mapping from the corresponding  stochastic process defined in the tangent bundle     $T{\cal P}$.

Thus  the behaviour of the original global
semigroup  is determined by  local evolution semigroups.
And these local semigroups are defined by  solutions of the local stochastic diffusion 
differential equations.
Therefore, it is possible to derive  the transformations of 
the global semigroup (\ref{eq4})  by studying transformations of the local
stochastic differential equations.

The global semigroup (\ref{eq4}) can be written (symbolically) as
\begin{eqnarray}
{\psi}_{t_b} (A_a({\mathbf x}),t_a)&=&{\rm E}\Bigl[\phi _0(\eta (t_b))
\exp \{\frac 1{\mu
^2\kappa }\int_{t_a}^{t_b}V[\eta (u)]du\}\Bigr]\nonumber\\
&=&\int_{\Omega _{-}}d\mu ^\eta (\omega )
\phi _0(\eta (t_b))\exp \{\ldots
\},
\label{2}
\end{eqnarray}
where ${\eta}_t({\bf x}) = {\eta}(t,\bf{x})$ is a stochastic process given 
on a manifold $\cal P$ (in our case -- on a manifold of the original gauge connections).
${\mu}^{\eta}$  is a measure generated by this process 
on the path space $\Omega _{-}=\{{\omega}_t \equiv
\omega (t,{\mathbf x}):\omega (t_a,{\mathbf x})=0,
\eta (t,{\mathbf x})=A_a({\mathbf x})+\omega (t,{\mathbf x})\}$. 

Such a  representation for the solution of the Kolmogorov equation given by the path integral over  the space of paths ${\Omega}_{-}$ is possible if on  charts of the  tangent bundle to the Hilbert manifold $\cal P$
there 
exist the self--adjoint, positive operators of trace class. Regarding each of these operators (usually defined by  the quadratic form) as the covariance operator of a 
 Gaussian measure one determines 
the Wiener processes and then the diffusion processes on charts of $T {\cal P}$. 
In fact, it is the case of \cite{Asorey}, where  the  manifold 
$\cal P$ was endowed with the regularized metric. 

The stochastic differential equation for the local components 
$\eta ^{(\alpha, i,x)}_t$ of the regularized stochastic process ${\eta}_t $ 
is as follows: \footnote{We present here only the diffusion part of the  equation.}
\begin{equation}
d\eta ^{(\alpha, i,x)}(t)=\mu \sqrt{\kappa }\;
{\mathfrak X}^{(\alpha, i,x)}_{\;\;\;\;\;\;\;\bar{M}}(\eta
(t))\;
dw^{
\bar{M}}(t)\,,
\label{3}
\end{equation}
where $w^{\bar{M}}(t)$ is a Wiener process  and  the ``matrix'' ${\mathfrak X}^{(\alpha, i,x)}_{\;\;\;\;\;\;\;\bar{M}}$ is derived  from the local equality: 
$\sum_{\bar{К}}
{\mathfrak X}_{\;\;\;\;\;\;\;\bar{K}}^{(\alpha, i,x)}{\mathfrak X}_
{\;\;\;\;\;\;\;\bar{K}}^{(\beta,j,y)}=G^{(\alpha,
i,x)\,(\beta,j,y)}$.
(The bar over indices indicates that these indices are to be taken as belonding to the Euclidean space.)
The regularized form of the metric leads to the matrices
${\mathfrak X}^{(\alpha, i,x)}_{\;\;\;\;\;\;\;\bar{M}}$ that include 
the factors $({\Roman 1}+{\triangle}_A/{\Lambda}^2)^{-k/2}$ with 
$k>3$ and even. The differential generator of the ``true'' stochastic process defined by the local stochastic differential equation, which has  the diffusion term with the same coefficient as in equations (\ref{3}) and also the corresponding drift term, is     the Laplace--Beltrami operator of the manifold with the regularized metric. 

In our formal approach we shall not deal with the regularization questions. It means that our original stochastic process is given on a Hilbert  manifold ${\cal P}$ with a plane Riemannian metric. 
That is,  we endow the Hilbert manifold $\cal P$ of the gauge fields in Sobolev class $H_k$ ($k>3$) with the Riemannian structure by choosing
the (weak) $L^2$  scalar product in its model space -- the Hilbert space ${\cal H}$. 
The stochastic process on  ${\cal P}$ is described by the local stochastic differential equation (\ref{3}) in which we set  
${\mathfrak X}^{\bar{A}}_{\;\;\bar{M}}={\delta}^{\bar{A}}_{\;\;\bar{M}}$. 

The process $w^{\bar{M}}(t)$ of equation (\ref{3})
is a ``cylindrical version'' of the Wiener process 
that can be constructed in the space of paths with the values in some  Hilbert space ${\cal H}_{-}$ in which there exists a  Gaussian measure.\footnote{The corresponding local stochastic differential equations for the processes with the values in ${\cal H}_{-}$ must  be  considered in a weak sense.}
 In the Hilbert space ${\cal H}$, in  case of using the $L^2$ weak scalar product, we can only determine the cylindrical measure, since the identity operator being the correlation operator  of the covariance given by this scalar product is not of the trace class.

The Wiener process $w_t$ with the values in ${\cal H}_{-}$ and which has the identity correlation operator in ${\cal H}$ is called  the canonical Wiener process. In appropriate It\^{o} calculus its properties in the Hilbert space ${\cal H}$ can be writen as follows:
\begin{eqnarray*}
 &&{\rm E}\,\Bigl(dw_t(f)\, dw_t(g)\Bigr)=dt\, (f,g)_{L^2}\\
&& {\rm E}\,\Bigl(dw_t(f)\Bigr)=0\,,
\end{eqnarray*}
or formally, as ${\rm E}\,\bigl(dw_t({\mathbf x})\, dw_t({\mathbf  y})\bigr)=dt\, {\delta}^3({\mathbf x}-{\mathbf y})$ and 
${\rm E}\,\bigl(dw_t({\mathbf x})\bigr)=0$.

Thus our stochastic differential equations (\ref{3})
can be used only for definition of the cylindrical measures 
on the space of the paths that have  their values in the Hilbert space $\cal H$.

Therefore we can consider the semigroup (\ref{eq4}) only as a formal expression presented by the superposition of  multiple integrals. 
These integrals are obtained from the cylindrical approximations of the local evolution semigroups and are equal to the integrals of the cylindrical functions. The integrals are taken over the corresponding cylindrical measures.
The rigorous definition of the semigroup (\ref{eq4}) acting in the space of functions given on the Hilbert space of distributions ${\cal H}_{-}$
should be based on the regularization of the metric and the renormalization  of the final expression.

The evolution semigroup (\ref{eq4}) acts in the Banach space of bounded and continuous functions (functionals) given on the space of gauge fields. The space of functions can be supplied by the scalar product 
\begin{equation}
\bigl(\psi\,,\psi \bigr)={\int}_{\cal P}\,
{\bar{\psi}}(A({\mathbf x}))\,{\psi}(A({\mathbf x}))\,\prod_{\mathbf x}
dA^{\alpha}_i(\mathbf x)\,,
\label{scalpr}
\end{equation}
which also has a formal sense, since the ``measure'' 
${\prod}_{\mathbf x} dA^{\alpha}_i({\mathbf x})$ is not defined as a Lebesgue measure.

It worth to notice that in case of exploiting the rigged Hilbert spaces for description of the evolution, one can
overcome  the problem of the definition of  the scalar product in the space of the  ``wave functions''. It can be done by using the Gaussian measure   instead of the formal measure of the scalar product (\ref{scalpr}). But in this case, because of  changing of  the ``natural''  normalization of the wave functions, the self-adjoint Hamilton operator has to include an additional term   \cite{Dalecky1}.

\section{Metric on $\cal P$ in bundle coordinates}

The gauge field $A$ defined on $M$ is the pull-back of a smooth
connection one-form given on a trivial principle fiber bundle $P(M,G)$  ($P=M\times G$).
 Here $M$ -- is a compact three--dimensional space ($S^3$ or $T^3$) and $G$ -- is a compact simple Lie group.
  We shall consider not a space of all smooth connections but 
its subspace $\cal P$ of the irreducible connections.
  For a  connected manifold $M$, the holonomy group  of each irreducible  connection coinsides with 
the group $G$.
 
 The gauge transformation group acts on a space of the
 irreducible connections. But this action is not free. To obtain a free action  one should factor 
 out the group of gauge transformations  by  the 
transformations taking their values in the centre of the group. 
 The resulting factor group, which we shall denote by $\cal G$, 
 consists of such  maps from $M$ to $G$ that belong to   the Sobolev class $H_{k+1}$ with $k>3$.

 From the slice theorem it follows \cite{Mitter} that $\cal P$ 
may be viewed as  a total space of a bundle of connections 
${\pi}: {\cal P}\to \cal M$ in which the base $\cal M={\cal P}/{\cal G}$ is   the space of gauge orbits.
 The space $\cal P$ can be locally presented  as  
${\pi}^{-1}({\cal U})\sim {\cal U}\times \cal G$, 
where ${\cal U}$  
 is a neigbourhood of a point $\pi(p)$ on the   base manifold 
$\cal M $.
Note that   existence of  such a priciple fiber bundle  was 
 proved in  \cite{Mitter,Parker,Soloviev,Paycha}.

 We shall define the coordinates on this principal  fiber 
bundle by extending 
an approach of \cite{Razumov} to the infinite dimensional case.
This approach was used in \cite{Storchak_2} when we studied the quantum motion 
of a scalar particle on a manifold with a given action of the compact
group Lie.
Notice that introduction of the coordinates by method of \cite{Razumov}  
is similar to what has been  done earlier in other papers 
concerning the gauge field quantization \cite{Christ, TudrCreutz,Falck}.

The fiber bundle coordinates of a point  $p\in \cal P$ will be determined 
by gauge fixing method.  We use the Coulomb gauge condition 
${\chi}^{\nu}(A)\equiv {\partial}^k A^{\nu}_k({\mathbf x})=0$, 
($\nu=1,\dots,N_{G}$) which is imposed on the gauge fields.
It means that original coordinates "$Q^A$'' 
(i.e.,   the gauge fields $A^{\alpha}_i({\mathbf x})$ in our case)
of a point $p\in \cal P$ 
can be expressed by means of the
coordinates of the corresponding point given on a gauge surface 
$\Sigma$  and a coordinates of a 
certain group element of a gauge group $\cal G$.

We shall use the following symbolical representation 
for the right action of the gauge  group 
on the space $\cal P$ of the irreducible connections: 
 \[
 \tilde A^{(\alpha,i,x)}=F^{(\alpha,i,x)}(A({\mathbf x}),
 g({\mathbf x}))\,.
 \]
An explicite form of the transformation is given by
\[
{\tilde A}^{\alpha}_i({\mathbf x})={\rho}^{\alpha}_{\beta}(g^{-1}(x))
{ A}^{\beta}_i({\mathbf x})+u^{\alpha}_{\mu}(g({\mathbf x}))
\frac{\partial g^{\mu}({\mathbf x})}
{\partial {\mathbf x}^i}\,,
\]
where matrix $u^{\alpha}_{\mu}(g({\mathbf x}))$
 is  analogous  to the matrix $u^{\alpha}_{\mu}(g)$
 which is  one of the auxilieries matrices of the compact Lie groups.
The left and right auxilieries matrices result from the differentiation of 
the group multiplication function (the multiplication table
given in the space of group parameters) by the group elements.

Recall that matrix $v^{\alpha}_{\beta}(g)$ is equal to
$\frac {\partial {\tilde\Phi}^{\alpha}(g,g_1)}
{\partial g^{\beta}_1}|_{g_1=e}$
and matrix ${u}^{\alpha}_{\beta}(g)$ is inverse to 
$v^{\alpha}_{\beta}(g)$:
$v^{\gamma}_{\beta}u^{\alpha}_{\gamma}=
 {\delta}^{\alpha}_{\beta}$. 
Likewise, the matrix ${\bar v}^{\alpha}_{\beta}(g)$ is defined by 
 ${\bar v}^{\alpha}_{\beta}(g)=\frac {\partial {\tilde\Phi}^
 {\alpha}(g_1,g)}
{\partial g^{\beta}_1}|_{g_1=e}$,
 and ${\bar u}^{\alpha}_{\beta}$ is its inverse matrix.  
 The matrix ${\rho}^{\alpha}_{\beta}(g)=\bar{u}^{\alpha}_{\nu}(g)\,
v^{\nu}_{\beta}(g)$ is  a matrix of the adjoint
 representation of a group $G$.
 An inverse matrix ${\rho}^{\alpha}_{\beta}(g^{-1})\equiv 
\bar{\rho}^{\alpha}_{\beta}(g)$
is defined by $\bar{\rho}^{\alpha}_{\tilde \beta}
{\rho}^{\tilde \beta}_{\mu}={\delta}^{\alpha}_{\mu}$.
\footnote{Matrix ${\rho}^{(\alpha,x)}_{\;\;\;(\beta,y)}=
{\rho}^{\alpha}_{\beta}(g({\mathbf x}))\,
{\delta}^3({\mathbf x}-{\mathbf y})$ will be 
the matrix of the adjoint representation 
of the gauge transformation group $\cal G$.}

 Let an arbitrary point $p$ with the coordinates $Q^A$ be given on a manifold $\cal P$.
And let the  action of a group $\cal G$ be given 
on this manifold.
With the help of the gauge surface $\Sigma$ (${\chi}^{\alpha}=0$), which has 
a transverse intersection with the orbits of the group action, 
we can determine such a group element $g \in {\cal G}$ that takes 
the point $p$ along the orbit  to the corresponding 
point on $ \Sigma $. The coordinates $g(Q)$ of this element $g$ 
can be obtained by solving  the following
equation:
\[
{\chi}^{\alpha}(\,F(Q,g^{-1}(Q))\,)
=0\,.   
\]
For the Coulomb gauge, this equation is as follows:
\[
\partial^i({\mathbf x})\left[\, {\rho}^{\alpha}_
{\;\beta}(g({\mathbf x}))\,
A^{\beta}_{\,i}({\mathbf x})-{\rho}^{\alpha}_{\;\nu}(g({\mathbf x}))\,
u^{\nu}_{\;\sigma}(g({\mathbf x}))\,\frac{\partial 
g^{\sigma}({\mathbf x})}
{\partial {\mathbf x}^i}\,\,\right]=0\,.
\]
That is, in order  to find 
the  element $g({\mathbf x})$, which takes 
 $A$ to the gauge surface ${\chi}^{\alpha}=0$,
 we must solve this equation provided  that
the gauge field $A({\mathbf x})$ is given.

The coordinates $Q^{\ast}$ of the corresponding point 
on a submanifold $\Sigma $ can be
obtained from the solution of the equation
$Q^{\ast}{}^A=F^A(Q,g^{-1}(Q))$, where $Q$ are the coordinates 
of the original point on the manifold ${\cal P}$.
In our case, the coordinates $Q^{\ast}$ are 
the dependent coordinates: the gauge connections 
$A^{\ast}{}^{\alpha}_i({\mathbf x})$ are subjected to the gauge condition  ${\chi}^{\alpha}(A^{\ast})=0$. 

Therefore  we have the   bijective correspondence
$A^{\alpha}_i({\mathbf x})\leftrightarrow 
 (A^{\ast}{}^{\alpha}_i({\mathbf x}),g^{\mu}({\mathbf x}))$
given by the gauge transformation
\[
A^{\alpha}_i({\mathbf x})={\rho}^{\alpha}_{\beta}(g^{-1}({\mathbf x}))
{ A^{\ast}}^{\beta}_i({\mathbf x})+u^{\alpha}_{\mu}(g({\mathbf
x}))\frac{\partial
g^{\mu}({\mathbf x})}
{\partial {\mathbf x}^i}\,.
\]
Of course, all this is valid if the equation for 
 $g^{\mu}({\mathbf x})$ has a unique
solution for a given  $A^{\ast}{}^{\alpha}_i({\mathbf x})$.

Notice that we are allowed to use 
the points of the surface $\Sigma$  for 
coordinatization of the total space $\cal P$ of the  
bundle of connections since there is 
a local isomorphism between the trivial principal bundle 
$\Sigma \times {\cal G} \to {\Sigma} $ and the principal bundle 
$P(\cal M, \cal G)$ (for the last bundle we have
locally  ${\pi}^{-1}({\cal U})\sim {\cal U}\times \cal G$).

Using the Coulomb gauge, we shall consider   the evolution  in a 
  sufficiently small neighbourhood of a point $p$ given on 
the original manifold. It will be also assumed that  the gauge surface has 
a transversal intersection with each gauge orbit. And hence we will assume the validity of the slice theorem in our case.

Changing $A^{\alpha}_i({\mathbf x})$ for 
$(A^{\ast}{}^{\alpha}_i({\mathbf x}),g^{\mu}({\mathbf x}))$, we chose new  coordinates on the manifold  $\cal P$
(the total space of the principal fiber bundle ${\pi}: {\cal P}\to \cal M$). Now our aim is to get a new coordinate representation of the original Riemannian metric
\[
ds^2=G_{(\alpha , i, x)(\beta,j,y)}\delta A^{(\alpha ,i,x)}
\delta A^{(\beta ,j,y)}\,,
\]
where
\[
G_{(\alpha , i, x)(\beta,j,y)}=G\biggl(\frac{\delta}
{\delta A^{\alpha}_i({\mathbf x})}\,,\,\frac{\delta}
{\delta A^{\beta}_j({\mathbf y})}\biggr)
=k_{\alpha \beta}\delta^{ij}{\delta}^3({\mathbf x}-{\mathbf y})\,.
\]
As in a finite  dimensional  case, 
it can be done if  we make a replacement of variables 
provided that the fact  of dependence of the gauge fields  $A^{\ast}$ will be  taken into account.
The ''vector fields'' transformation formula  is  a strightforward
generalization of the corresponding formula from the finite
dimensional case:
\begin{eqnarray}
&&\frac{\delta}{\delta A^{(\alpha,i,x)}}=
\check F^{(\mu, k,u)}_{\;\;\;\;\;\;(\alpha,i,x)}\,
N^{(\nu,p,v)}_{\;\;\;\;\;\;(\mu, k,u)}(A^{\ast})\,
\frac{\delta}{\delta A^{\ast}{}^{(\nu,p,v)}}\nonumber\\
&&\;\;\;\;+\check F^{(\epsilon,
m,z)}_{\;\;\;\;\;\;(\alpha,i,x)}\,
{\chi}^{(\mu,v)}_{\;\;\;(\epsilon, m,z)}(A^{\ast})\,
\bigl(\Phi ^{-1}\bigr)^{(\beta,u)}_{\;\;\;(\mu,v)}(A^{\ast})\,
{\bar v}^{(\sigma,p)}_{\;\;(\beta,u)}(g)\,
\frac{\delta}{\delta g^{(\sigma,p)}}\,,
\label{5}
\end{eqnarray}
where we have denoted by $\check F$ the matrix 
which is inverse to the matrix  
$F^{(\mu, k,u)}_{\;\;\;\;\;\;(\alpha,i,x)}$  defined as follows:
\[
F^{(\alpha,i,x)}_{\;\;\;\;\;\;(\beta,j,y)}[A,g]=
\frac{\delta {\tilde A}^{(\alpha,i,x)}}
{\delta A^{(\beta,j,y)}}=
{\rho}^{\alpha}_{\beta}(g^{-1}({\mathbf x}))\, \delta^i_{\,j}\,
{\delta}^3  ({\mathbf x}-{\mathbf y})\,.
\]
It satisfies the relation:
\[
F^{(\alpha,i,x)}_{\;\;\;\;\;\;(\beta, j,y)}\,
\check F^{(\beta, j,y)}_{\;\;\;\;\;\;(\epsilon, k,z)}=
{\delta}^{\alpha}_{\,\epsilon}\,{\delta}^i_{\,k}\,
{\delta}^3({\mathbf x}-{\mathbf z})\,.
\]
Note that matrix $F^{(\mu, k,u)}_{\;\;\;\;\;\;(\alpha,i,x)}$ 
acts in the tangent space to the manifold  $\cal P$.

In formula (\ref{5}), by $N^{(\nu,p,v)}_{\;\;\;\;\;\;(\mu, k,u)}$,
which is equal to
\[
N^{(\alpha,i,x)}_{\;\;\;\;\;\;(\beta,j,y)}=
{\delta}^{(\alpha,i,x)}_{\;\;\;\;\;\;(\beta,j,y)}
-K^{(\alpha,i,x)}_{\;\;\;\;\;\;(\mu,z)}
\bigl({\Phi}^{-1}\bigr)^{(\mu,z)}_{\,\,\;\;(\nu,u)}
{\chi}^{(\nu,u)}_{\;\;\;\;(\beta,j,y)}\,,
\]
we have denoted the projection operator 
onto the subspace which is orthogonal to the Killing 
vector field $K_{(\alpha,y)}$.
For our metric $G_{(\alpha , i, x)(\beta,j,y)}$, 
the Killing vector field $K_{(\alpha,y)}$  is
\[
K_{(\alpha,y)}=K^{(\mu, i, x)}_{\;\;\;\;\;\;(\alpha,y)}\frac{\delta}
{\delta A^{(\mu ,i,x)}}\,,
\]
where
\[
K^{(\mu, i, x)}_{\;\;\;\;\;\;(\alpha,y)}(A)=
\left[\left({\delta}^{\;\mu}_{\alpha}{\partial}^i({\mathbf x})
+c^{\mu}_{\tilde \nu \alpha}A^{\tilde \nu i}({\mathbf x})
\right){\delta}^3 ({\mathbf x}-{\mathbf y})\right]
\equiv \left[{\mathcal D}^{\mu i}_{\;\;\alpha}(A({\mathbf x}))
\,{\delta}^3({\mathbf x}-{\mathbf y})\right]
\]
(here ${\partial}_i({\mathbf x})$ is a partial derivative 
with respect to $x^i$).
These vector fields are obtained by taking the functional derivative 
of $F^{(\alpha,i,x)}$ with respect to $g({\mathbf x})$ and then setting $g$ to an identity.
Our Killing vector fields 
\[
K_{(\alpha,y)}=  \left(-{\delta}^{\mu}_{\;\alpha}\;
{\partial}_i({\mathbf y})
+c^{\mu}_
{\tilde \nu \alpha}A^{\tilde \nu}_i({\mathbf y})\right)
\frac{\delta}
{\delta A^{\mu}_i({\mathbf y})}\equiv
{\tilde {\mathcal D}}^{\mu }_{\;\alpha\, i}(A({\mathbf y}))\frac{\delta}
{\delta A^{\mu}_i({\mathbf y})}
\]
(there is no an integration with respect to $\mathbf y$)
act in the space of functions that depend on the gauge connections.

The Faddeev--Popov matrix $\Phi $ is defined as follows:
\[
{\Phi}^{(\nu,y)}_{\;\;(\mu,z)}\bigl[A\bigr]=
K^{(\alpha,i,x)}_{\;\;\;\;(\mu,z)}\;{\chi}^{(\nu,y)}_
{\;\;(\alpha,i,x)}\,.
\]
For the Coulomb gauge, we have
\[
{\chi}^{(\nu, y)}_{\;\;\;\;(\alpha,i,x)}=
{\delta}^{\nu}_{\;\alpha}
\left[\,{\partial }_i({\mathbf y})\;{\delta}^3({\mathbf y}-
{\mathbf x})\,\right]\,.
\]
Therefore, the matrix $\Phi $ (restricted to the gauge surface) is equal to
\[
{\Phi}^{(\nu,y)}_{\;\;\;\;(\mu,z)}[A^{\ast}]=
\left[\bigl({\delta}^{\nu}_{\mu}\;{\partial}^2({\mathbf y})
+c^{\nu}_{\sigma
\mu}{A^{\ast}}{}^{\sigma}_i({\mathbf y})\;
{\partial}^i({\mathbf y})\,\bigr)\;
{\delta}^3({\mathbf y}-{\mathbf z})\right]
\]
or
\[
{\Phi}^{(\nu,y)}_{\;\;\;\;(\mu,z)}[A^{\ast}]=
\left[\bigl(\,{\cal D}\,[A^{\ast}]\cdot\partial \,
\bigr)^{\nu}_{\mu}({\mathbf y})\;{\delta}^3({\mathbf y}-{\mathbf
z})\,\right]\,.
\]
An inverse matrix ${\Phi}^{-1}$ can be determined  by the equation
\[
{\Phi}^{(\nu,y)}_{\;\;\;\;(\mu,z)}\;
({\Phi}^{-1}){}^{(\mu,z)}_{\;\;\;\;(\sigma,u)}({\mathbf y},
{\mathbf u})=
{\delta}^{\nu}_{\sigma}\,{\delta}^3({\mathbf y}-{\mathbf u})\,
\]
That is, it is the Green function for the Faddeev--Popov operator:
\[
\left[\partial_i({\mathbf y}){\cal
D}^{\nu\,i}_{\mu}[A({\mathbf y})]\,\right]\,
({\Phi}^{-1}){}^{(\mu,y)}_{\;\;\;\;(\sigma,u)}
({\mathbf y},{\mathbf u})=
{\delta}^{\nu}_{\sigma}\,{\delta}^3({\mathbf y}-{\mathbf u})\,.
\]
(The boundary conditions of this operator depend on a concrete choice
of a base manifold $M$.)
By a second group of variables, the Green function ${\Phi}^{-1}$  satisfies  the following equation:
\[
\left[-\tilde{\cal D}^{\sigma\,i}_{\lambda}[A({\mathbf z})]\,
\partial_i({\mathbf z})\,\right]\,({\Phi}^{-1}){}^{(\mu,y)}_
{\;\;\;\;(\sigma,z)}
({\mathbf y},{\mathbf z})=
{\delta}^{\mu}_{\lambda}\,{\delta}^3({\mathbf y}-{\mathbf z})\,.
\]
Notice that in the formula (\ref{5}), the matrix ${\Phi}^{-1}$,  
as well as the other terms of
the projector  $N$, is given on the gauge surface $ \Sigma$.

In definition of the transformed metric, besides the operator $N$,  we shall also use   another 
projection operator
$\bigl(P_{\bot}\bigr)^
{(\alpha,k,x)}_{\;\;\;\;\;\;(\beta,m,y)}$.
This operator performs the projection onto the tangent plane to the submanifold $\Sigma$ and  can be written symbolically as
\[
\bigl(P_{\bot}\bigr)^{(\alpha,k,x)}_{\;\;\;\;\;\;(\beta,m,y)}=
{\delta}^{\alpha}_{\,\beta}\left[{\delta}^k_{\,m}+
{\partial}_m\frac{1}{(-{\partial}^2)}
{\partial}^k\right]
{\delta}^3({\mathbf y}-{\mathbf x})\,.
\]
Its definition and properties (together with the properties of the
projection operator $N$) are given in  Appendix.

In new coordinates, the metric ${\tilde G}_{\cal A\cal B}
(A^{\ast},g)$ of the manifold $\cal P$ is presented by the matrix: 
\begin{equation}
\displaystyle
{\tilde G}_{\cal A\cal B}(A^{\ast},g)=
\left(
\begin{array}{cc}
{\tilde G}_{(\alpha,i,x)\,(\beta, j,y)}
& {\tilde G}_{(\alpha,i,x)\,(\gamma,y)}
\\ {\tilde G}_{(\gamma,y)\,(\alpha,i,x)}
& {\tilde G}_{(\alpha,x)\,(\beta,y)}
\end{array}
\right)\,.
\label{6}
\end{equation}
It has the following elements with respect to the basis 
$(\frac{\delta}{\delta A^{\ast}}\,,
\frac{\delta}{\delta g})$:
\begin{eqnarray*}
{\tilde G}_{(\alpha,i,x)\,(\beta, j,y)}&=&
G_{(\tilde{\alpha},m,\tilde x)\,(\tilde\beta,n,\tilde y)}
\left(P_{\bot}\right)^{(\tilde{\alpha},m,\tilde x)}_
{\;\;\;\;\;\;(\alpha,i,x)}\;
\left(P_{\bot}\right)^{(\tilde\beta,n,\tilde y)}_
{\;\;\;\;\;\;(\beta, j,y)}\;\nonumber\\
&&=k_{\tilde \alpha \beta}\,\delta_{i n}
\left(P_{\bot}\right)^{(\tilde \alpha,n,x)}_
{\;\;\;\;\;\;(\alpha,j,y)}\,.
\end{eqnarray*}
The off-diagonal elements of this metric are
\begin{eqnarray*}
&&{\tilde G}_{(\alpha,i,x)\,(\gamma,y)}=
G_{(\tilde\alpha,m,\tilde x)\,(\tilde \beta, n,\tilde y)}
\left(P_{\bot}\right)^{(\tilde \beta, n,\tilde y)}_
{\;\;\;\;\;\;(\alpha,i,x)}\;
K^{(\tilde\alpha,m,\tilde x)}_{\;\;\;(\mu,u)}\,{\bar u}^
{(\mu,u)}_{\;(\gamma,y)}\nonumber\\
&&\;\;\;\;\;\;\;\;\;\;\;\;\;\;\;\;\;
=k_{\tilde \alpha \tilde \beta}\,{\delta}_{mn}\,
c^{\tilde \alpha}_{\;\sigma \mu}\,A^{\ast}{}^{\sigma m}(v)\,
\left(P_{\bot}\right)^{(\tilde \beta, n,y)}_
{\;\;\;\;\;\;(\alpha,i,x)}\;{\bar u}^
{\mu}_{\,\gamma}(g({\mathbf y}))\,.
\end{eqnarray*}
 The element of ${\tilde G}_{(\alpha,x)\,(\beta,y)}$ of the transformed metric  is equal to
\begin{eqnarray*}
&&{\tilde G}_{(\alpha,x)\,(\beta,y)}={\gamma}_{(\mu,\tilde x)\,(
\nu,\tilde y)}\,
{\bar u}^{(\mu,\tilde x)}_{\;(\alpha,x)}\,
{\bar u}^{(\nu,\tilde y)}_{\;(\beta,y)}\nonumber\\
&&\;\;\;\;\;\;\;
=k_{\epsilon \sigma}\,{\delta}^{k l}\,
\left[{\tilde {\cal D}}^{\epsilon }_{\mu \,k}
[A^{\ast}({\mathbf x})]\,
{\cal D}^{\sigma }_{\nu \,l}[A^{\ast}({\mathbf y})]\,
\delta ^3({\mathbf x}-{\mathbf y})\right]\,
{\bar u}^{\mu}_{\;\alpha}(g({\mathbf x}))
{\bar u}^{\nu}_{\;\beta}(g({\mathbf y}))\,.
\end{eqnarray*}
By analogy with  the finite dimensional case, 
the orbit metric $\gamma _{(\mu,x)( \nu,y)}$ can be defined as
\[
\gamma _{(\mu,x)( \nu,y)}=K^{(\alpha,i,z)}_{\;\;\;\;\;\;(\mu,x)}
G_{(\alpha,i,z)(\beta,j,u)}K^{(\beta,j,u)}_{\;\;\;\;\;\;( \nu,y)}\,.
\]
Hence, 
\[
\gamma _{(\mu,x)( \nu,y)}=\int d^3u\,d^3v\,k_{\varphi
\alpha}\,{\delta}^{kl}\,
\delta ^3({\mathbf u}-{\mathbf v})
\left[{\cal D}^{\varphi }_{\mu \,k}({\mathbf u})
\delta ^3({\mathbf u}-{\mathbf x})\right]
\left[{\cal D}^{\alpha }_{\nu \,l}({\mathbf v})
\delta ^3({\mathbf v}-{\mathbf y})\right].
\]
Metric $\gamma $  may be explicitely written   in the following form:
\[
\gamma _{(\mu,x)( \nu,y)}=k_{\varphi \alpha}{\delta}^{kl}
\left[\bigl(-\delta
^{\varphi}_{\,\mu}\;{\partial}_k({\mathbf x})+c^{\varphi}_{\sigma
\mu}A^{\sigma}_k({\mathbf x})\bigr)\bigl(\delta
^{\alpha}_{\,\nu}{\partial}_l({\mathbf y})+c^{\alpha}_{\kappa
\nu}A^{\kappa}_l({\mathbf y})\bigr){\delta}^3({\mathbf x}-{\mathbf
y})\right].
\]

Matrix ${\tilde G}^{\cal A\cal B}(A^{\ast},g)$,
\begin{equation}
\displaystyle
{\tilde G}^{\cal A\cal B}(A^{\ast},g)=
\left(
\begin{array}{cc}
{\tilde G}^{(\alpha,i,x)\,(\beta, j,y)}
& {\tilde G}^{(\alpha,i,x)\,(\sigma,y)}
\\ {\tilde G}^{(\sigma,y)\,(\alpha,i,x)}
& {\tilde G}^{(\sigma ,x)\,(\epsilon ,y)}
\end{array}
\right),
\label{7}
\end{equation}
 is  inverse to the matrix ${\tilde G}_{\cal A\cal B}$.
The matrix elements of  ${\tilde G}^{\cal A \cal B}$  are given by
\[
{\tilde G}^{(\alpha,i,x)\,(\beta, j,y)}=
G^{(\tilde\alpha,m,\tilde x)\,(\tilde\beta, n,\tilde y)}\,
N^{(\alpha,i,x)}_
{\;\;\;\;\;\;(\tilde\alpha,m,\tilde x)}\,
N^{(\beta, j,y)}_{\;\;\;\;\;\;(\tilde\beta, n,\tilde y)}\,;
\]
\[
{\tilde G}^{(\alpha,i,x)\,(\sigma,y)}=
G^{(\tilde{\alpha},n,\tilde x)\,(\tilde\beta, m,\tilde y)}\,
N^{(\alpha,i,x)}_{\;\;\;\;\;\;(\tilde{\alpha},n,\tilde x)}\,
{\Lambda}^{(\nu, y)}_{\;\;(\tilde\beta, m,\tilde y)}\,
{\bar v}^{\sigma}_{\,\nu}(g({\mathbf y}))\,.
\]
(In the previous formula there is no an integration with respect to $\mathbf y$).

A new term   $\Lambda $  is defined as follows:
\[
{\Lambda}^{(\nu, u)}_{\;\;(\tilde\beta, m,\tilde y)}=
{\chi}^{(\mu ,\tilde z)}_{\;\;(\tilde\beta, m,\tilde y)}\,
\left(\Phi ^{-1}\right)^{(\nu, u)}_
{\;\;(\mu,\tilde z)}\,.
\]
Its explicit representation  is 
\[
{\Lambda}^{(\nu, u)}_{\;\;(\tilde\beta, m,\tilde y)}=
(-1)\,\left[\,\partial _m (\tilde {\mathbf y})
\left(\Phi ^{-1}\right)^{(\nu, u)}_
{\;\;(\tilde \beta,\tilde y)}({\mathbf u},\tilde {\mathbf y})\,\right]\,.
\]
The matrix element ${\tilde G}^{(\sigma ,x)\,(\epsilon ,y)}$
 of the matrix ${\tilde G}^{\cal A \cal B}$  may be written as
\[
{\tilde G}^{(\sigma ,x)\,(\epsilon ,y)}=
G^{(\tilde{\alpha},n,\tilde x)\,(\tilde\beta, m,\tilde y)}\,
{\Lambda}^{(\nu, x)}_{\;\;(\tilde{\alpha},n,\tilde x)}\,
{\Lambda}^{(\mu, y)}_{\;\;(\tilde\beta, m,\tilde y)}\,
{\bar v}^{\sigma}_{\,\nu}(g({\mathbf x}))\,
{\bar v}^{\epsilon}_{\,\mu}(g({\mathbf y}))\,
\]
(there is no here an integration with respect to $x$ and $y$).

Notice that matrix elements of ${\tilde G}_{\cal A \cal B}$
and ${\tilde G}^{\cal A \cal B}$ are restricted 
to the gauge surface, that is, they  depend on $A^{\ast}$.

The matrices (\ref{6}) and (\ref{7}) are pseudo inverse to each other:
\begin{eqnarray*}
\displaystyle
{\tilde G}^{\cal A\cal B}{\tilde G}_{\cal B\cal C}=\left(
\begin{array}{cc}
(P_{\perp})^{(\mu, k, u)}_{\;\;\;\;(\nu,m,v)}  & 0\\
0 & {\delta}^{(\alpha,x)}_{\;\;(\beta,y)}
\end{array}
\right)\,.
\end{eqnarray*}
The determinant of the matrix (\ref{6}) is equal to
\begin{eqnarray*}
&&(\det {\tilde G}_{\cal A\cal B})=
\det G_{AB}(Q^{\ast})\det {\gamma}_{\alpha \beta}(Q^{\ast})
(\det {\chi}{\chi }^{\top})^{-1}(Q^{\ast})
(\det {\bar u}^{\mu}_{\nu}(a))^2\nonumber\\
&&\,\,\,\,\,\,\,\,\,\,\times(\det 
{\Phi}^{\alpha}_{\beta} (Q^{\ast}))^2
\det {}^{'} (P_{\perp})^C_B(Q^{\ast})\,.
\end{eqnarray*}
An explicit form of this determinant is given 
by the following formula:
\[
\det {\tilde G}_{\cal A \cal B}= \det k_{\alpha \beta}\;
{\det}^{-1}(-{\partial}^2)\,{\det}^2\,( {\bar u}^{\mu}_{\,\nu})\,\;
{\det}^2 \, (\,\partial\cdot {\cal D}^{\alpha}_{\beta}
[A^{\ast}]\,)\,{\det}\, {}^{'} P_{\perp}\,, 
\]
where 
\[
\det \, {}^{'}
\bigl(P_{\bot}\bigr)^{(\alpha,k,x)}_{\;\;\;\;\;\;(\beta,m,y)}
=
\det \, {}^{'}
\left({\delta}^k_{\,m}+
{\partial}_m\frac{1}{(-{\partial}^2)}
{\partial}^k\right)\,.
\]

\section{Transformation of the stochastic process\\ 
and the semigroup}

In the previous section,  we have introduced
 new local coordinates on the manifold  $\cal P$ of 
the gauge connections.
These coordinates have been  obtained by means of  
transformation  of the original variables.

A similar local transformation  can be done for 
the stochastic process
$\eta (t)$ and we come to  another stochastic 
process $\zeta (t)$.
The local components 
$(A^{\ast}_t({\mathbf x}),g_t({\mathbf x}))$
of the process $\zeta (t)$ can be
expressed  through the components of the 
process $\eta (t)$ by the  gauge transformation formulae: 
\[
{\eta}^{(\alpha, i,x)}(t)=F^{(\alpha,i,x)}(A^{\ast}(t),g(t))\,.
\]

The local processes are given on  charts 
of the manifold $\cal P$. These processes are   correspond stochastically to each other in  general domains of the charts intersections.
By applying the method of \cite{Dalecky}, we can then construct the global stochastic process in  that part of the manifold 
$\cal P$, where we are allowed to consider the Coulomb gauge as a true gauge. 

To  transform the global semigroup (\ref{eq4}), we,  first of all, make   the transformations of the local semigroups (\ref{locsemi}).  
 Our transformations of the local stochastic processes are the phase space transformation 
of the stochastic processes. It is well known that these transformations conserve the probabilities.
Therefore, the transformations will leave invariant the path integral measures in our local semigroups.

Measures of our path integrals are generated by the stochastic processes defined by the
 solutions of the stochastic differential equations. 
After performing the transition to the process $\zeta (t)$, we come to a new semigroup represented in the form
 of mathematical expectation of initial function taken with respect to the measure generated by the 
transformed process.

We are interested in  the stochastic differential equations for the  local components 
$(A^{\ast}_t({\mathbf x}),g_t({\mathbf x}))$ of the process
$\zeta (t)$. 
Let us   recall  how the similar stochastic differential
 equations have been obtained \cite{Storchak_2} in a finite dimensional case.
In that case we considered the transformation of  the stochastic process $\eta (t)$ 
 given  on a compact Riemannian manifold $\cal P$. 

Let $Q^{\ast}{}^A(t)$ be the components  of the 
process ${\zeta}^A (t)=(Q^{\ast}{}^A(t), g^{\alpha}(t))$. 
Since there is a  bijective correspondence  between $Q^A$ and
$(Q^{\ast}{}^A, g^{\alpha})$,  we can express the coordinates $Q^{\ast}{}^A$ in terms of the original coordinates $Q^A$:
\[
Q^{*}{}^A=F^A(Q,g^{-1}(Q))\,.
\]
The same relation can be written for the components of the corresponding stochastic processes.  
Performing the stochastic differentiation of  the  random variable $Q^{*}{}^A(t)$ 
with the It\^{o} formula, we get 
   the following equation:
\begin{equation}
dQ^{*}{}^A(t)=\frac{\partial Q^{*}{}^A}{\partial
Q^E}d{\eta}^E(t)
+\frac12\frac{{\partial}^2Q^{*}{}^A}{\partial Q^E 
\partial Q^C}<d{\eta}^E(t)d{\eta}^C(t)>.
\label{sd2}
\end{equation}  
If  we substitute   (\ref{3}) (which in a general case has the corresponding drift term) for $d{\eta}^A(t)$ in (\ref{sd2}) 
and  replace then the stochastic variable $Q^A(t)$  by 
$Q^{*}{}^A(t)$ and $g^{\alpha}(t)$  
according to the formula ${\eta}^A(t)=
F^A(Q^{*}{}^B(t),g^{\alpha}(t))$, 
we shall obtain the drift and diffussion terms  of the stochastic differential equation of the local 
process $Q^{*}{}^A(t)$. Such a transformation performed with   account of the invariance of the original metric, leads us to the following stochastic differential equation for the local process $Q_t^{*}$:
\begin{eqnarray}
&&dQ_t^{*}{}^A=
 \frac12{\mu}^2\kappa\,\left [ 
 -G^{LM}(Q^{\ast}_t)\,\Gamma^E_{LM}(\,F(Q^{\ast}_t,{\rm e})\,)
 N^A_E(Q^{\ast}_t)\,\right.
  \nonumber\\
&&\,\,\,+G^{LM}(Q^{\ast}_t)\,\left.
 N^A_{LM}(Q^{\ast}_t)\right]\,dt
 +{\mu}
 \sqrt{\kappa} N^A_C(Q^{\ast}_t)\, 
 {\mathfrak X}^C_{\bar M}(Q^{\ast}_t)\,dw^{\bar M}(t)
\label{sde1}
\end{eqnarray}
where $\Gamma $ - are the Christoffel symbols and by $N^A_{LM}$ we have denoted the partial derivative of $N^A_L$ with respect to $Q^{*}{}^M$.

 The stochastic differential equation for 
 the  components of the process $\zeta (t)$ that are given on a group manifold $\cal G$ of the principal fiber bundle $P(\cal M,\cal G)$
  can be  obtained  by using the similar transformations.

The  stochastic differential equation (\ref{sde1}) has been rewritten \cite{Storchak_2} in order to be appropriate for performing the reduction in the path integral. The drift term of this equation has been separated into two parts.
 The first part  is responsible for the stochastic
 movement on the gauge surface $\Sigma $. The second  one, 
 denoted by $j_{\Roman 2}^A(Q^{*})$,  is the  projection of 
the mean curvature of the group orbit onto the tangent plane to $\Sigma $. The obtained stochastic differential equation was 
  \begin{equation}
 dQ^{*}{}^{\small A}(t)={\mu}^2\kappa
 \biggl(-\frac12
 G^{EM}N^C_EN^B_M\,{}^H{\Gamma}^A_{CB}+
 j^{\small A}_{\Roman 1}+j^{\small A}_{\Roman 2}\biggr)dt +\mu\sqrt{\kappa}
 N^A_C\tilde{\mathfrak X}^C_{\bar M}dw^{\bar M}\,,
 \label{sd8}
\end{equation}
 where $j_{\Roman 2}^A$ is \footnote{In  \cite{Storchak_2},
 this term was written with a wrong sign.}
 \[
  j_{\Roman
  2}^A(Q^{*})=-\frac12\, {\mu}^2\kappa G^{EU}N^A_EN^D_U
  \left[{\gamma}^{\alpha\beta}G_{CD}
  ({\tilde{\nabla}}_{K_{\alpha}}
  K_{\beta})^C\right](Q^{*})\,.
  \]
 The $j_{\Roman 2}^A$  may also be  written as follows:
 \[
  j_{\Roman
  2}^A(Q^{*})=-\frac12\,{\mu}^2\kappa N^A_C\,
  \left[{\gamma}^{\alpha\beta}
  ({\tilde{\nabla}}_{K_{\alpha}}
  K_{\beta})^C\right](Q^{*})\,.
  \]
Note that the first part of the drift term of the equation (\ref{sd8}) includes \cite{Storchak_2} the Christoffel coefficient ${}^H{\Gamma}$ of the ``horizontal'' connection  and the mean curvature 
$j_{\Roman 1}$ of the orbit space ${\cal M}$.

  Let us consider the particular case when 
  the original metric is plane, that is, $G_{AB}={\delta}_{AB}$.
  This case is important for us because in the diffusion problem of  the 
  Yang--Mills fields we also have a plane  metric given  on the manifold of the gauge connections.
For the plane metric   the equation (\ref{sde1}) can be written as follows:
\begin{equation}
dQ_t^{*}{}^A=
 \frac12{\mu}^2\kappa\,\left [ 
 -N^A_E\,G^{MC}\,
 K^E_{\;\beta C}\,{\Lambda}^{\beta}_M\,\right]dt
+{\mu}
 \sqrt{\kappa} N^A_C\, 
 {\mathfrak X}^C_{\bar M}\,dw^{\bar M}(t)\,.
\label{sde22}
\end{equation}  
  (The terms on the right-hand side of this equation 
  depend on $Q^{\ast}_t$.)

  As in general case, we rewrite the  coefficient of the drift term in    the stochastic differential equation (\ref{sde22}). 
Now we present the mean curvature  $j_{\Roman 2}$ of this  term as
\[
j^A_{\Roman 2}=\frac12\,({\mu}^2 \kappa)\,N^A_{\,C}\,
G^{BC}\, K^P_{\,\sigma\, B}\, {\mathscr A}^{\sigma}_{\,P}\,,
\]
where we have denoted by  ${\mathscr A}^{\sigma}_{\,P}$ a new quantity
\[
{\mathscr A}^{\sigma}_P(Q^{\ast})={\gamma}^{\sigma\,\mu}(Q^{\ast})
\,K^R_{\mu}(Q^{\ast})\,G_{RP}(Q^{\ast})\,.
\]
It is related to the natural connection 
one--form ${\mathscr A}^{\sigma}_i(x)$ that exists in the considered principal fiber bundle $\pi :{\cal P} \to {\cal P}/{\cal G}$, namely,
\[
{\mathscr A}^{\sigma}_i(x)={\mathscr A}^{\sigma}_P
(Q^{\ast}(x))\,Q^{\ast}{}^P_i(x)\,.
\]
Here,  $x$  are  independent coordinates on the orbit space ${\cal M}={\cal P}/{\cal G}$, 
and $Q^{\ast}{}^P_i(x)$ are the partial derivatives (with respect to $x^i$) of the  functions $Q^*(x)$ that "resolve" the gauge 
condition ${\chi}^{\alpha}(Q^*(x))=0$.

The connection ${\mathscr A}^{\sigma}_i(x)$ was called the mechanical connection
in the reduction problems of the classical mechanic \cite{AbrMarsd}. 
Its analog  in  Yang--Mills quantization is called the Coulomb connection.  

Using this connection, we rewrite equation (\ref{sde22}) in the following form
\begin{equation}
dQ^{\ast}_t{}^A=\frac12\,{\mu}^2\kappa\,
\Bigl(N^A_{\,E}\,G^{B\,E}\, K^{L}_{\,\epsilon\,B}\,(\,{\Lambda}^
{\epsilon}_{\, L}-{\mathscr A}^{\epsilon}_{\,L}\,)\,dt+
j_{\Roman 2}^A\Bigr) dt+{\mu}{\kappa}^{1/2}\,
N^A_{\,C}{\cal X}^C_{\bar M}dw^{\bar M}\,.
\label{sde23}
\end{equation}

What we have done 
with the stochastic process and its stochastic differential equation in a finite dimensional case 
can also be carried out with the Yang--Mills stochastic process.

We may transform the local process ${\eta}^{(\alpha,i,x)}_t$ defined by the equation (\ref{3}) to the local 
process ${\zeta}^{(\alpha,i,x)}_t$ 
by making use of the It\^o formula\footnote{ 
This formula for the stochastic differential was proved  in \cite{Asorey} where  the diffusion process 
in the space of the Yang--Mills connections was considered.}
written in terms of the functional derivatives.
As in the finite dimensional case, the process ${\zeta}_t$
has  two components: $(A^{\ast}{}^{(\beta,j,x)}_t,
g^{(\mu,x)}_t)$, that is, 
$A^{\ast}_t{}^{(\beta,j,x)}\equiv A^{\ast}_t{}^{\beta\, j}
({\mathbf x})$
 and $g^{(\mu,x)}_t\equiv g^{\mu}_t({\mathbf x})$.\footnote{The existence of the process $g_t$  can be proved 
by using an approach which is similar to that one developed for 
the stochastic process given on the gauge group \cite{Gaveau_2}.}

Performing the necessary transformation, we come to the stochastic
differential equation which looks like 
the analogous equation obtained in the finite dimensional case.
  Therefore, we shall use the equation (\ref{sde23})  assuming now that the values of this equation have 
the generalized indices, that is, the indices have  
the discrete and the continuous components.

Thus, the equation (\ref{sde23}) will be regarded as the symbolical representation of the ``true'' stochastic differential equation, the explicit form of which can be easily obtained  
by integrating  over the repeated continuous indices.
In the sequel,  we shall also employ an analogous symbolical representations for the other stochastic differential equations.
 The  main terms of these equations will be also presented in the  explicit form.

As for the representation of the terms of the equation (\ref{sde23}) written for  diffusion of the Yang--Mills fields, the connection ${\mathscr A}^{\alpha}_{\,B}$  is given by the following expression:
\[
{\mathscr A}^{(\alpha,x)}_{\;\;\;(\beta,j,y)}=
\left[{\cal D}^{{\varphi}}_{\;\mu j}(A^{\ast}({\mathbf y}))
{\gamma}^{(\alpha,x)\,(\mu , y)}\right]\,k_{\varphi \beta}\,.
\]
In the stochastic differential equations adapted to 
the Yang--Mills diffusion, $K^C_{\,\alpha B}$  denotes the functional
derivative of $K^{(\epsilon , m,z)}_ {\;\;(\alpha,x)}$ with respect to $A^{(\beta,j,y)}$:
\[
K^{(\epsilon , m,z)}_{(\alpha,x)(\beta,j,y)}\equiv
\frac{\delta}{\delta A^{(\beta,j,y)}}\,
K^{(\epsilon , m,z)}_{\;\;\;\;(\alpha,x)} =
{\delta}^m_{\,j}\,c^{\epsilon}_{\,\beta \alpha}\,
{\delta}^3 ({\mathbf z}-{\mathbf x})\,
{\delta}^3 ({\mathbf z}-{\mathbf y})\,.
\]
Summing on the generalized index $\alpha$ in 
the expression ${\mathscr A}^{\alpha}_{\,С}\,K^C_{\,\alpha B}$ of the eq.(\ref{sde23}), we get
\[
k_{{\varphi} \epsilon}\,\,\left.
c^{\epsilon}_{\,\beta \alpha}\left[
{{\cal D}}^{\varphi}_
{\;\mu j }(A^{\ast}({\mathbf z}))\,
\,\gamma ^{( \alpha,y)( \mu,z)}\right]\right |_{z=y}\,.
\]
This  expression includes an "inverse matrix" 
$\gamma ^{( \alpha,y)( \mu,z)}$ for    matrix $\gamma _{(\mu,x)( \nu,y)}$. 
It can be defined by the following equation:
\[
\gamma _{(\mu,x)( \nu,y)}\;
\gamma ^{( \nu,y)( \sigma,z)}={\delta}^{( \sigma,z)}_
{\;(\mu,x)}\equiv{\delta}^{\sigma}_
{\,\mu}\,{\delta}^3({\mathbf z}-{\mathbf x})\,.
\]
Performing the integration with respect to  $\mathbf y$ on the left-hand side of this equation, we get
\[
k_{\varphi \alpha}\,{\delta}^{kl}\,
{\tilde{\cal D}}^{\varphi }_{\mu \,k}(A^{\ast}(\mathbf x))\,
{{\cal D}}^{\alpha }_{\nu \,l}(A^{\ast}(\mathbf x))
\,\gamma ^{( \nu,x)( \sigma,z)}={\delta}^{\sigma}_
{\,\mu}\,{\delta}^3(\mathbf z-\mathbf x)\,.
\]
Thus, $\gamma ^{( \nu,x)( \sigma,z)}$ is the Green function of the operator $({\tilde{\cal D}}\,{{\cal D}})_{\mu \nu}$.
In some of our expressions it  will be denoted by 
$\gamma ^{( \nu,x)( \sigma,z)}[A^{\ast}]$ 
in order that to stress its implicit dependence on $A^{\ast}$.

Finally, the $j_{\Roman 2}^A$ -- term of the stochastic equation can be rewritten as follows:
\[
j^{(\alpha ,i,x)}_{\,{\Roman 2}}=-1/2\,({\mu}^2\, \kappa) 
\int dz\,
N^{(\alpha ,i,x)}_{\;\;\;\;\;\;(\epsilon,m,z)}\,
\left.
c^{\epsilon}_{\,\sigma \nu}\left[
{\cal D}^{\sigma m }_{\,\mu}(A^{\ast}(y))\,
\,\gamma ^{( \mu,y)( \nu,z)}\right]\right |_{y=z}\,.
\]

Note also that  
if  we make use of  an explicit expression of the projector
$N$  and perform the corresponding integrations (``the summation'' over 
repeated generalized indices in eq.(\ref{sde23}) )                     we shall come to a rather long expression. By this reason we 
shall keep this projector in its symbolical form 
during the course of our transformations.

The stochastic differential equation for the group--valued component 
$g_t$ of the process ${\zeta}^{(\alpha,i,x)}_t$
can be obtained in the same way as it was
done for the component $A^{\ast}_t$. 
The  equation for the $g^{(\mu,x)}_t\equiv g_t^{\mu}({\mathbf x})$  coinsides by its form with  the corresponding stochastic differential
equation for the group component of the finite dimensional case:
\begin{eqnarray}
&&dg^{\alpha}_t=-\frac12{\mu}^2\kappa\biggl[G^{RP}{\Lambda}^{\sigma}_R
{\Lambda}^{\beta}_BK^B_{\sigma P}
{\bar v}^{\alpha}_{\beta}
-G^{CA}N^M_C\frac{\delta}
{\delta A^*{}^M}
\biggl({\Lambda}^{\beta}_A\biggr)
{\bar v}^{\alpha}_{\beta}
\nonumber\\
&&\,\,\,\,\,\,\,\,\,\,
-G^{MB}{\Lambda}^{\epsilon}_M
{\Lambda}^{\beta}_B\,{\bar v}^{\nu}_{\epsilon}\,
\frac{\delta}
{\delta g^{\nu}}
\bigl({\bar v}^{\alpha}_{\beta}\bigr) 
\biggr]dt+\mu\sqrt{\kappa}\,
{\bar v}^{\alpha}_{\beta}{\Lambda}^{\beta}_B\,
\tilde{\mathfrak X}^B_{\bar M}\,dw^{\bar M}.
\label{sd10}
\end{eqnarray}
We shall also make use of this equation in the  diffusion 
of the Yang--Mills fields taking into account 
the remarks that have already been done on the stochastic differential equation for the process $A^{\ast}_t$.

Notice that the equation (\ref{sd10}) can also be  written as follows:
\begin{eqnarray}
&&dg^{\alpha}_t=-\frac12{\mu}^2\kappa\biggl[
\biggl(2\,G^{AC}{\Lambda}^{\beta}_B
{\Lambda}^{\epsilon}_C\,K^B_{\epsilon C}
-G^{AC}\,{\Lambda}^{\beta}_B\,{\Lambda}^{\epsilon}_A\,
{\Lambda}^{\varphi}_C\,K^B_{\varphi M}\,K^M_{\epsilon}
\biggr)\,{\bar v}^{\alpha}_{\beta}
\nonumber\\
&&\,\,\,\,\,\,\,\,\,\,
-G^{MB}{\Lambda}^{\epsilon}_M
{\Lambda}^{\beta}_B{\bar v}^{\nu}_{\epsilon}
\frac{\delta}
{\delta g^{\nu}}
\bigl({\bar v}^{\alpha}_{\beta}\bigr) 
\biggr]dt+\mu\sqrt{\kappa}\,
{\bar v}^{\alpha}_{\beta}\,{\Lambda}^{\beta}_B\,
\tilde{\mathfrak X}^B_{\bar M}\,dw^{\bar M}.
\label{sd11}
\end{eqnarray}

The obtained stochastic differential equations, (\ref{sde23}) and (\ref{sd11}), 
determine the local process ${\zeta}^{(\alpha,i,x)}_t$ and, consequently, a new semigroup
${\tilde U}_{\zeta}$.
Since the process ${\zeta}^{(\alpha,i,x)}_t$ was obtained from the process ${\eta}^{(\alpha,i,x)}_t$ 
by  the phase space transformation of the stochastic processes, 
we have
\[
 {\tilde U}_{\eta}(s,t) \phi _0 (A)=
 {\rm E}_{s,A}[\phi _0 (\eta (t))]=
 {\rm E}_{s,(A^{\ast},g)}[{\tilde{\phi}} _0 (\zeta (t))]
 ,\,\,\,
 s\leq t,\,\,\,\eta (s)=A({\mathbf x}),
 \]
$\zeta (s)=(A^{\ast}({\mathbf x}),g({\mathbf x}))$  and  $A({\mathbf x})$ is related to the initial 
value of the process ${\zeta} (t)$ by the  gauge transformation: 
$A({\mathbf x})=F(A^{\ast}({\mathbf x}),g({\mathbf x}))$.

Hence, we have the following equality for the local semigroups:
 \[
 {\tilde U}_{\eta}(s,t) \phi _0 (A)=
 {\tilde U}_{\zeta}(s,t) {\tilde{\phi}}_0 (A^{\ast},g).
 \]
The corresponding  superposition of the local semigroups ${\tilde U}_{\zeta}$ leads, as in \cite{Dalecky}, to the global semigroup 
${\tilde U}_{\zeta}$ of the process $\zeta (t)$:
\begin{equation}
\psi _{t_b}(A_a({\mathbf x}),t_a)=
{\lim}_q {\tilde U}_{\zeta}(t_a,t_1)\cdot\ldots\cdot
{\tilde U}_{\zeta}
(t_{n-1},t_b)\, 
{\tilde \phi} _0(A^*_a, g_a)\,,
\label{sd21}
\end{equation}
This global semigroup which  is the  transformation of the 
original semigroup (\ref{eq4}) determines the transformed path 
integral. 

 The path integral defined by (\ref{sd21}) will be symbolically written  as 
\[
{\psi}_{t_b} (A_a({\mathbf x}),t_a)={\rm E}\Bigl[\tilde{\phi
}_0({\xi}_{\Sigma}(t_b),g(t_b))\exp
\{\frac 1{\mu ^2\kappa }\int_{t_a}^{t_b}
\tilde{V}({\xi}_{\Sigma}(u))du\}\Bigr],
\]
where  ${\xi}_{\Sigma} (t_a)=A^{*}_a({\mathbf x})$, 
$g(t_a)=g_a({\mathbf x})$.
Using the It\^o formula (in its functional form), we get 
the following  symbolical representation of the differential generator (the Hamilton operator) of the semigroup (\ref{sd21}) related to the process $\zeta (t)$:
\begin{eqnarray*}
&&\frac12{\mu}^2\kappa\left(G^{CD}N^P_CN^B_D
\frac{{\delta}^2}{\delta A^{*}{}^P
\delta A^{*}{}^B}
+N^P_{\,E}\,G^{B\,E}\, K^{L}_{\,\epsilon\,B}\,(\,{\Lambda}^
{\epsilon}_{\, L}-{\mathscr A}^{\epsilon}_{\,L}\,)
\frac{\delta}{\delta A^{*}{}^P}\right.
\nonumber\\
&&+N^E_{\,C}\,
G^{BC}\, K^P_{\,\sigma\, B}\, {\mathscr A}^{\sigma}_{\,P}
\frac{\delta}{\delta A^{*}{}^E}
+G^{AB}
{\Lambda}^{\alpha}_A{\Lambda}^{\beta}_B
{\bar L}_{\alpha}{\bar L}_{\beta}
-G^{RP}
{\Lambda}^{\sigma}_R {\Lambda}^{\alpha}_B
K^B_{\sigma P} {\bar L}_{\alpha}\nonumber\\
&&+G^{CA}N^M_C
\frac{\delta}{\delta A^{*}{}^M}
\left({\Lambda}^{\alpha}_A\right)
{\bar L}_{\alpha}
\left. +\,2\,G^{BC}N^P_C{\Lambda}^{\alpha}_B
{\bar L}_{\alpha}\frac{\delta}
{\delta A^{*}{}^P}\right)+\frac{1}
{{\mu}^2\kappa }V\,,
\end{eqnarray*}
where by ${\bar L}_{\alpha}$ we denote $ {\bar L}_{\alpha}=
{\bar v}^{\mu}_{\alpha}(g({\mathbf x}))\frac{\delta}
{\delta g^{\mu}({\mathbf x})}$.
Note also that all terms of the obtained differential generator 
(except for the ${\bar L}_{\alpha}$) depend on 
$A^{\ast}({\mathbf x})$.

An explicit form of the differential generator may be easily obtained from its symbolical expression as a result of the corresponding integration over the repeated continuous indices.

\section{Path integral measure factorization}

Since in our investigation we assume that the stochastic 
processes can be defined by the method of  \cite{Dalecky}, 
we shall make use of the approach to the factorization of the path
integral measure  that has been developed in \cite{Storchak_2}.
This approach is based on the application of 
the nonlinear filtering equation from the stochastic 
processes theory.

Taking into account the  Markov property of the process $\zeta (t)$, 
we present each
 local semigroup  ${\tilde U}_{\zeta}$  of the   equation (\ref{sd21})  as follows:
\begin{equation}
{\tilde U}_{\zeta}(s,t) {\tilde
\phi} (A^{*}_0,g_0)=
{\rm E}
\Bigl[{\rm E}\bigl[\tilde{\phi }(A^{*}(t),g(t))\mid
(
{\cal F}_{A^{*}})_{s}^{t}\bigr]\Bigr]\,,
\label{25}
\end{equation}
According to  the optimal nonlinear filtering theory
{\cite{Lipster,Pugachev}, the  conditional mathematical expectation 
of $\widetilde{\phi }$ given the sub-$\sigma$-algebra $
{\cal F}_{A^{*}}$,
\[
\hat{\widetilde{\phi }}(A^{*}(t))\equiv 
{\rm E}\Bigl[\tilde{\phi }(A^{*}(t),g(t))\mid (
{\cal F}_{A^{*}})_{s}^{t}\Bigr],
\]
satisfies  the certain stochastic differential equation 
which is called  the nonlinear filtering equation.
Solving this equation, we get an information on the stochastic 
process $g_t$  provided that we observe the process $A^{*}_t$.

It is important that the coefficients of the stochastic differential
equations of the processes $g_t$  and $A^*_t$ depend  on the stochastic variables in the following  way:
\begin{eqnarray*}
\left\{
\begin{array}{l}
\displaystyle
dA^{\ast}_t=a_1(A^{\ast}_t,g_t,t)\,dt+X_1(A^{\ast}_t,t)\,dw_t\\
dg_t=a(A^{\ast}_t,g_t,t)\,dt+X(A^{\ast}_t,g_t,t)\,dw_t\,.
\end{array}
\right.
\end{eqnarray*} 
 It can therefore be possible to derive the stochastic nonlinear filtering
equation {\cite{Lipster,Pugachev} from the stochastic 
differential equations of the processes $A^{\ast}_t$ and $g_t$:
\begin{eqnarray*}
&&d\hat{\widetilde{\phi }}(A^{*}_t)={\rm E}\left[\tilde{\phi }_t+\tilde{\phi }_g\,a+\frac12\,{\tilde{\phi }}_{gg}
\,
(XX^{\top})\left|({\cal F}_{A^{*}})_{s}^{t}
\right. \right]dt\\
&&\,\,\,+\,{\rm E}\left[\tilde{\phi }\,\{a_1-\hat{a}_1\}+ 
\tilde{\phi }_g\,(XX_{1}^{\top})\left|
(
{\cal F}_{A^{*}})_{s}^{t}
\right.\right](X_{1}X_{1}^{\top})^{-1}
(dA^{\ast}_t-\hat{a}_1dt),
\end{eqnarray*}
where�� $\hat{a}_1={\rm E}[{a}_1(A^{\ast}_t,g_t,t) 
|({\cal F}_{A^{*}})_{s}^{t}]$ and $\tilde{\phi }_t$  is the partial derivative  of the $\tilde{\phi }$ 
with respect to $t$, and  $\tilde{\phi }_g$ -- the corresponding 
derivative with respect to $g$.

Note that in our case  the drift coefficient $a_1$ of the stochastic  differential equation  (\ref{sde23}) does not depend on 
the group--valued process $g_t$.

From the symmetry of our problem it follows that the nonlinear filtering stochastic differential equation  is a linear equation. 
It symbolical form looks like an analogous equation of a finite 
dimensional case \cite{Storchak_2}:
\begin{eqnarray}
&&d\hat{\widetilde{\phi }}(A^{*}_t)=-\frac12{\mu}^2\kappa
\Bigl(
G^{RP}{\Lambda}^{\sigma}_R{\Lambda}^{\beta}_B
K^B_{P\sigma} - G^{CA}N^M_C
\frac{\delta}{\delta A^{*}{}^M}
({\Lambda}^{\beta}_A)\Bigr)
\nonumber\\
&&\times{\rm E}\left
[\bar{L}_\beta \tilde{\phi }\left|
({\cal F}_{A^{*}})_{s}^t\right.\right]dt
+\frac12\,{\mu}^2\kappa \,
G^{CB}{\Lambda}^{\nu}_C{\Lambda}^{\kappa}_B
{\rm E}\left
[\bar{L}_\nu\bar{L}_{\kappa} \tilde{\phi }
\left|({\cal F}_{A^{*}})_{s}^t\right.\right]dt
\nonumber\\
&&+{\mu}\sqrt{\kappa}\,{\Lambda}^{\beta}_C\,{\Pi}^C_K\,
{\tilde{\cal X}}^K_{\bar M}\,{\rm E}
\left[\bar{L}_\beta \,\tilde{\phi }\left|
({\cal F}_{A^{*}})_{s}^t\right.\right]\,dw_t^{\bar M}\,,
\label{26}
\end{eqnarray} 
where   by ${\Pi}$ we denote the projection operator onto the 
"horisontal subspace". This  operator is defined  by the equation:  ${\Pi}^A_B={\delta}^A_B-K^A_{\alpha}\,
 {\gamma}^{\alpha\,\beta}\,K^C_{\beta}\,G_{CB}$ and  has 
the following properties: ${\Pi}^A_B\,K^B_{\alpha}=0$,  ${\Pi}^C_E\,N^E_B={\Pi}^C_B$, ${\Pi}^B_E\,N^T_B=N^T_E$.

Using the properties of the conditional mathematical expectations 
for the Markov processes, we could change 
the equation (\ref{26}) for some solvable  equation. 
But, it could be done if we were able to factorise the variables $A^{*}$ and $g$ in our  functional $\tilde{\phi }$.
 In a finite dimensional case,  we have used  the Peter--Weyl theorem.  It allowed us to develop  the function given on 
a compact group in series over the irreducible repesentations of this group. 

In order to apply the similar approach to the problem considered 
in the present paper, we shall make  use of the  irreducible finite dimensional unitary 
 representation constructed in \cite{Rossi}. This representation is given by the functional of the local  group elements $g({\mathbf x})$. For a compact group $G$, this functional depends on 
the values assumed by $g({\mathbf x})$ in a finite number of points.

The matrices of the representation are written as follows:
\[
Y^{\lambda}(g(x))={\exp}\bigl(i\,(J_{\mu}){}^{\lambda}\,g^{\mu}
({\mathbf x})\bigr),
\]
where $(J_{\mu}){}^{\lambda}$ are the infinitesimal generators of the representation: 
$
\bar{L}_\mu Y_{pq}^\lambda (g)=\sum_{q^{\prime
}}(J_\mu )_{pq^{\prime
}}^\lambda Y_{q^{\prime }q}^\lambda (g).
$ 

We shall assume that it is  possible to develop the  functional 
$\tilde{\phi }$ in a series over these  irreducible representation:
\[
\tilde{\phi }(A^{*},g)=\sum_{\lambda
,p,q}r_{pq}^\lambda
(A^{*})Y_{pq}^\lambda (g)\,. 
\]
Making use of such  a representation for $\tilde{\phi }$ in the conditional mathematical expectation, we obtain
\begin{equation}
{\rm E}\bigl[\tilde{\phi }(A^{*}(t),g(t))\mid
({\cal F})
_{A^{*}})_{s}^t\bigr]=
\sum_{\lambda ,p,q}r_{pq}^\lambda (A^{*}(t))\,
\hat{Y}_{pq}^\lambda(A^{*}(t))\,,
\label{N1}
\end{equation}
where    $\hat{Y}_{pq}^\lambda(A^{*}(t))={\rm E}\bigl[Y_{pq}^\lambda
(g(t))\mid ({\cal F}_{A^{*}})_{s}^t\bigr]$.
Since    $r_{pq}^\lambda $ only depends  on $A^{\ast}_t$, 
we have taken it  out of the conditional expectation.

Notice that  the conditional mathematical expectation $\hat{Y}_{pq}^\lambda (A^{*}(t))$ besides $A_t^{*}$
depends also on initial values of the stochastic processes
(i.e.,  it depends on $A^{*}_0(\mathbf x)=A_s^{*}(\mathbf x)$ 
and $g_0^{\alpha}(\mathbf x)=g_s^{\alpha}(\mathbf x)\,$). 

Using the relation (\ref{N1}) in eq.(\ref{26}), we get 
the following equation for $\hat{Y}_{pq}^\lambda $:
\begin{eqnarray}
&&d\hat{Y}_{pq}^\lambda (A^{*}(t))=
\nonumber\\
&&-\frac12{\mu}^2\kappa\left\{\Bigl[
G^{RP}{\Lambda}^{\sigma}_R{\Lambda}^{\mu}_B
K^B_{P\sigma}-\Bigl. G^{CP}N^M_C
\frac{\delta}{\delta A^{*}{}^M}({\Lambda}^{\beta}_P)
 (J_\beta)_{pq^{\prime}}^\lambda \hat{Y}_{q^{\prime }q}^\lambda
(A^{*}(t))\Bigr]\right.
\nonumber\\
&&-\left. G^{CB}{\Lambda}^{\alpha}_C {\Lambda}^{\nu}_B
\,\,(J_\alpha)_{pq^{\prime }}^\lambda 
(J_\nu)_{q^{\prime }q^{\prime \prime }}^\lambda
\hat{Y}_{q^{\prime \prime
}q}^\lambda (A^{*}(t))\right\}dt
\nonumber\\
&&+\mu\sqrt{\kappa}{\Lambda}^{\nu}_C{\Pi}^C_K
(J_\nu )_{pq^{\prime }}^\lambda \hat{Y}_{q^{\prime}q}^\lambda
(A^{*}(t)){\tilde{\cal  X}}_{\bar{M}}^K(A^{*}(t))dw^{\bar{M}}(t).
\label{28}
\end{eqnarray}
This stochastic differential equation is a matrix 
linear equation. Its solution may be written \cite{Dalmult}
via the multiplicative stochastic integral:
\footnote{The multiplicative stochastic integral is defined as  
a limit of the sequence of time--ordered multipliers that have been 
obtained as a result of breaking the time interval $[s,t]$. 
The direction of the arrow means that the order is taken from $s$ to $t$.}
\begin{equation}
\hat{Y}_{pq}^\lambda(A^{*}(t))=
(\overleftarrow{\exp })_{pn}^\lambda
(A^{*}(t),t,s)\,{\rm E}\bigl[Y_{nq}^\lambda
(g_s(\mathbf x))\mid ({\cal F}_{A^{*}})_{s}^t\bigr]\,,
\label{29}
\end{equation} 
where
\begin{eqnarray}
&&(\overleftarrow{\exp })_{pn}^\lambda
(A^{*}(t),t,s)=\overleftarrow{\exp}
\int_{s}^t\Bigl\{\frac 12{\mu}^2\kappa
\Bigl[\bar{\gamma }^{\sigma
\nu }(A^{*}(u))\,(J_\sigma
)_{pr}^\lambda (J_\nu )_{rn}^\lambda 
\nonumber\\
&&-\biggl(
G^{RP}{\Lambda}^{\sigma}_R\,{\Lambda}^{\beta}_B
\,K^B_{P\sigma}-G^{CA}N^M_C \frac{\delta}{\delta A^{*}{}^M}
({\Lambda}^{\beta}_A)\biggr)\,(J_\beta)_{pn}^\lambda \Bigr]du
\nonumber\\
&&+\mu\sqrt{\kappa}\,{\Lambda}^{\beta}_C\,(J_\beta)_{pn}^\lambda 
\,{\Pi}^C_K\,{\tilde {\cal X}}^K_{\bar M}\,dw^{\bar M}(u)\,
\Bigr\}.
\label{30}
\end{eqnarray}
Notice that for the conditional expectation taken at the initial values we have the equalities:
\[
{\rm E}\left[\,Y_{nq}^\lambda (g_s(\mathbf x))\left | ({\cal
F}_{A^{*}})_{s}^t\,\right.\right]
=Y_{nq}^\lambda (g_s(\mathbf x))=Y_{nq}^\lambda (g_0(\mathbf x)).
\]
Thus, the local semigroup (\ref{25}) may be written as follows:
\begin{equation}
{\tilde U}_{\zeta}(s,t) {\tilde\phi} (A^{*}_0(\mathbf x),
g_0(\mathbf x))=\sum_{\lambda ,p,q,q^{\prime }}{\rm E}
\left[\,r_{pq}^\lambda (A_t^{*})\,
(\overleftarrow{\exp })_{pq^{\prime }}^\lambda 
(A_t^{*},t,s)\,\right] Y_{q^{\prime}q}^\lambda (g_0(\mathbf x))\,.
\label{31}
\end{equation}

If we make the  similar transformations in all local semigroups 
that determine  the global semigroup (\ref{sd21}),  
we come to  the following representation: 
\begin{equation}
{\psi}_{t_b}(p_a,t_a)
=\sum_{\lambda ,p,q,q^{\prime }}{\rm E}
\left[\,
 c_{pq}^\lambda ({\xi}_{\Sigma}(t_b))
(\overleftarrow{\exp })_{pq^{\prime }}^\lambda 
({\xi}_{\Sigma}(t),t_b,t_a)\right]
Y_{q^{\prime}q}^\lambda (g_{a})
\label{32}
\end{equation}
$({\xi}_{\Sigma}(t_a)={\pi}|_{\Sigma}\circ p_a$=
$A^{\ast}_a({\mathbf x})$\,),  
where ${\xi}_{\Sigma}(t)$ is a global stochastic process given on
 a submanifold $\Sigma$. The local components of the process 
 ${\xi}_{\Sigma}(t)$ are the stochastic processes $A^{\ast}_t$.

 Let us  assume that 
there exists a fundamental solution  $G_{\cal P}(p_b,t_b;p_a,t_a)$ 
of our  equation (\ref{1}), that is,
\begin{equation}
{\psi}_{t_b} (p_a,t_a)=\int G_{\cal
P}(p_b,t_b;p_a,t_a)\phi_0(p_b)dv_{\cal P}(p_b)\,,
\label{fac5}
\end{equation}
where $dv_{\cal P}$ is a ``volume measure'' on $\cal P$.
 It is then possible to transform (\ref{32}) into the relation between the kernels (the Green functions) of the corresponding 
semigroups, provided that we choose the delta--function as an initial function of our original semigroup.

The  Green function $G_{\cal P}$, related to the Feynmann  propagator, may be expressed as a  sum of the 
 the matrix Green function $G^{\lambda}_{mn}$ multiplied by 
 the matrix of the irreducible representation $Y_{mn}^\lambda $.
Such a representation give us the  relation between two path
integrals. One of the path integrals describes the stochastic 
evolution on the manifold  ${\cal P}$ of the gauge connections, 
another one -- represents the evolution given on the gauge
 surface $\Sigma $ defined by the dependent variables $A^{\ast}$.

 The kernel of the evolution semigroup (the Green function $G^\lambda_{mn}$)  given by the mathematical expectation standing under the sign of the sum in (\ref{32})   can be symbolically written as follows:
 \begin{eqnarray}
&&G^{\lambda}_{mn}(\pi _{\Sigma}(p_b),t_b;
\pi _{\Sigma}(p_a),t_a)=
\nonumber\\
&&{\tilde {\rm E}}_{{\xi _{\Sigma} (t_a)=
\pi _{\Sigma}(p_a)}\atop  
{\xi _{\Sigma}(t_b)=\pi _{\Sigma}(p_b)}}
\left[(\overleftarrow{\exp })_{mn}^\lambda 
(\xi _{\Sigma}(t),t_b,t_a)
\exp \left\{\frac 1{\mu ^2\kappa }
\int_{t_a}^{t_b}\tilde{V}(\xi _{\Sigma}(u))
du\right\}\right]
\nonumber\\
&&=\int\limits_{{\xi _{\Sigma}(t_a)=
\pi _{\Sigma}(p_a)}\atop  
{\xi _{\Sigma}(t_b)=\pi _{\Sigma}(p_b)}} 
d{\mu}^{\xi _{\Sigma}}
\exp \left\{\frac 1{\mu ^2\kappa
}\int_{t_a}^{t_b}\tilde{V}(\xi _{\Sigma}(u))
du\right\}
\nonumber\\
&&\times\overleftarrow{\exp}
\int_{t_a}^{t_b}\Bigl\{\frac 12{\mu}^2\kappa
\Bigl[{\gamma }^{\sigma
\nu }(\xi _{\Sigma}(u))\,(J_\sigma
)_{mr}^\lambda (J_\nu )_{rn}^\lambda 
\Bigr.\Bigr.
\nonumber\\
&&+
G^{RP}{\Lambda}^{\sigma}_R{\Lambda}^{\beta}_B\,
K^B_{P\sigma}-G^{CP}N^M_C
\frac{\delta}{\delta A^{*}{}^M}
({\Lambda}^{\beta}_P)\biggr)
\,\,(J_\beta)_{mn}^\lambda \Bigr]du
\nonumber\\
&&+\Bigl.\mu\sqrt{\kappa}\,{\Lambda}^{\beta}_C\,
(J_\beta)_{mn}^\lambda 
{\Pi}^C_K\,{\tilde {\cal X}}^K_{\bar M}\,dw^{\bar M}(u)\Bigr\}\,,
\label{39}
\end{eqnarray}
($\pi _{\Sigma}(p_b)=A^{\ast}_b({\mathbf x})$ and 
$\pi _{\Sigma}(p_a)=A^{\ast}_a({\mathbf x})$). 
 In eq.(\ref{39}), the path integral measure ${\mu}^{\xi _{\Sigma}}$  is generated  by the process ${\xi}_{\Sigma}(t)$.
 The local components of this process are the solutions of the stochastic differential equation (\ref{sde23}).

The differential generator of the semigroup given by eq.(\ref{39}) is 
 \begin{eqnarray}
&&\frac12\mu ^2\kappa \left\{\left[
G^{CD}N^A_CN^B_D\frac{{\delta}^2}{\delta
A^{*}{}^A\delta
A^{*}{}^B}
\right.\right.
\nonumber\\
&&+\left.\,N^A_{\,E}\,G^{B\,E}\,
K^{L}_{\,\epsilon\,B}\,{\Lambda}^
{\epsilon}_{\, L}\,
\frac{\delta}{\delta A^*{}^A}
\right](I^\lambda )_{pq}
+2N^A_CG^{CP}{\Lambda}^{\alpha}_P
(J_\alpha )_{pq}^\lambda
\frac{\delta}{\delta A^{*}{}^A}
\nonumber\\
&&-\left(G^{RP}{\Lambda
}^{\sigma}_R{\Lambda}^{\alpha}_B\,K^{B}_{P\sigma}-G^{
CA}N^M_C\frac{\delta}{\delta
A^*{}^M}({\Lambda}^{\alpha}_A)\right)
(J_\alpha )_{pq}^\lambda 
\nonumber\\
&&+\biggl. G^{SB}{\Lambda}^{\alpha}_B
{\Lambda}^{\sigma}_S\,
(J_\alpha)_{pq^{\prime }}^\lambda 
(J_\sigma)_{q^{\prime }q}^\lambda \biggr\}+\frac{1}
{{\mu}^2\kappa }V\,.
\label{op2}
\end{eqnarray} 
($(I^\lambda )_{pq}$ -- is an identity matrix.)
 
This operator  acts in the space of sections ${\Gamma}(\Sigma,V^*)$ of the corresponding 
 covector bundle which is associated to the principal fiber bundle
$\pi: \Sigma \times {\cal G} \rightarrow \Sigma$.
As in a finite dimensional case  \cite{Storchak_2}, it can be shown  that the scalar product (\ref{scalpr}) is transformed to the formal scalar product in the space  ${\Gamma}(\Sigma,V^*)$ given by 
\[
 (\psi _n,\psi _m)=\!\!
 \int \langle \psi _n,\psi _m
 {\rangle}_{V^{\ast}_{\lambda}}
 \det{\Phi}^{\alpha}_{\beta}
 \prod_{\alpha =1}^{N_{ G}}
 \delta({\chi}^{\alpha}(A^{*}))
  \,\prod_{\mathbf x} d\,A^{\ast}({\mathbf x})\,,
    \]
where ${\chi}^{\alpha}(A^{*})={\partial}^k A^{*}_k{}^{\alpha}(\mathbf x)$
and $\det{\Phi}^{\alpha}_{\beta}=
{\det} \, (\,\partial\cdot {\cal D}^{\alpha}_{\beta}
[A^{\ast}]\,)$  is the determinant of the Faddeev--Popov operator 
in the Coulomb gauge.

In order to inverse the ``global'' relation between the Green functions which can be derived from (\ref{32}) it needs to assume the existence of the  partition of unity for the corresponding local covering of the manifold $\cal P$. Provided that this assumption takes place, one may recover  the necessary integral relation between the ``global'' Green functions from the integral  
relations obtained by inversion of the local Green functions given on charts of this partition. The latter may be done by using the
 orthogonal relations of the matrix elements of the 
 irreducible representation  from \cite{Rossi} and properties of the transition functions of  the principal fiber bundle. 
As a result one gets the following integral relation:
\begin{equation}
G^{\lambda}_{mn}({\pi}_{\Sigma} (p_b),t_b;
{\pi}_{\Sigma} (p_a),t_a)=
\int _{\cal G}G_{\cal P}(p_b\,g,t_b;
p_a,t_a) 
Y_{nm}^\lambda (g({\mathbf x}))\,d\mu (g({\mathbf x}))\,,
\label{38}
\end{equation}
where $d\mu (g({\mathbf x}))$ is the  normalized  invariant volume element given on a group $\cal G$ and by ``$p_b\,g$''  we have denoted the  gauge transformed boundary ``point''  $A_b({\mathbf x})$.
The obtained integral relation is defined (formally) on that  domain of the manifold $ \cal P$ where using of the Coulomb gauge does 
not violate the slice theorem. A similar  integral  relation between Green functions was  obtained in \cite{Rossi, Teitelboim}.

In order to get the path integral representation for the kernel of the evolution semigroup acting in the space  of the scalar functions given on  a submanifold $\Sigma $ one must set $\lambda =0$ in eq.(\ref{39}). This converts the multiplicative stochastic integral into  the identity matrix. And the differential generator of the obtained  semigroup  will be the diagonal part of the Hamiltonian (\ref{op2}).

  The  path integral measure of the $\lambda =0$   case
 is generated by the stochastic process defined by 
 the local stochastic differential equations (\ref{sde23}). But the equations of such a form have an "extra" term $j_{\Roman 2}$. 
 The differential generator of the process  determined 
 by the same local stochastic differential equations, but without 
these extra terms, would  
be  a Laplace--Beltrami operator for the  submanifold $\Sigma $.
The diffusion on $\Sigma $ governed by the Laplace--Beltrami  operator is directly related to the diffusion on the gauge orbit space.

Therefore we  make use of the Girsanov transformation in order 
to get a necessary description of the evolution on a gauge surface 
$\Sigma $. The Girsanov transformation changes  the path integral measure ${\mu}^{\xi}\equiv {\mu}^1$ generated by the process  ${\xi}_t$ whose local stochastic differential equations are
\[
dA^{\ast}_t{}^C=\frac12\,{\mu}^2\kappa\,
\Bigl(N^C_{\,E}\,G^{B\,E}\, K^{L}_{\,\epsilon\,B}\,(\,{\Lambda}^
{\epsilon}_{\, L}-{\mathscr A}^{\epsilon}_{\,L}\,)\,dt+
j_{\Roman 2}^A\Bigr) dt+{\mu}{\kappa}^{1/2}\,
N^C_{\,D}\,{\cal X}^D_{\bar M}\,dw_t^{\bar M}
\] 
for  the path integral measure ${\mu}^2\equiv {\mu}^{\tilde{\xi}_{\Sigma}}$ related to the stochastic process ${\tilde{\xi}_{\Sigma}}(t)$ with the local equations
\[
dA^{\ast}_t{}^C=\frac12\,{\mu}^2\kappa\,
N^C_{\,E}\,G^{B\,E}\, K^{L}_{\,\epsilon\,B}\,(\,{\Lambda}^
{\epsilon}_{\, L}-{\mathscr A}^{\epsilon}_{\,L}\,)\,dt
 +{\mu}{\kappa}^{1/2}\,
N^C_{\,D}\,{\cal X}^D_{\bar M}\,dw_t^{\bar M}\,.
\]
The Jacobian of the transformation is given by the following general formula: 
\[
\frac{d {\mu}^1}{d{\mu}^2}=\exp\Bigl\{-\frac12\int_{t_a}^{t_b}
[A^{-1}(b-a)]^2 dt+\int_{t_a}^{t_b}(A^{-1}(b-a),dw_t)\Bigr\}\,.
\]
As in \cite{Storchak_2}, in our case we will have
\begin{eqnarray*}
&&\frac{d {\mu}^1}{d{\mu}^2}({\tilde{\xi}_{\Sigma}}(t))=
\exp\Bigl\{
\int_{t_a}^{t}\Bigl[-\frac12
{\mu}^2\kappa\,
(P_{\bot})^L_A \,G^{H}_{LK}\,(P_{\bot})^K_E\,
j^A_{\Roman 2}\,j^E_{\Roman 2}\,\Bigr]
dt+\nonumber\\
&&\;\;\;\;\;\;\;\;\;\;\;\;\;\;\;\;\;\;\;\;\;\;\;+
{\mu}\,{\kappa}^{1/2}\,
G^{H}_{LK}\,(P_{\bot})^L_A\, j^A_{\Roman 2}\,
{\cal X}^K_{\bar M}\,dw_t^{\bar M}\,\Bigr]\Bigr\}\,,
\end{eqnarray*}
since
\[
(b-a)^A=-j^A_{\Roman 2}=\frac12{{\gamma}^{\nu\, \sigma}}
[{\nabla}_{K_{\nu}}K_{\sigma}]^C\,N^A_C\,.
\]
The term of the Jacobian with the stochastic integral 
may be transformed for the following expression:
\[
-\frac12\,{\mu}\,{\kappa}^{1/2}\,\int_{t_a}^{t_b}\,
G^H_{\,PC}\,{\gamma}^{\nu\sigma}[{\nabla}_{K_{\nu}}K_{\sigma}]^C\,
{\cal X}^P_{\bar M}\,dw_t^{\bar M}\,.
\]
The Jacobian of the Girsanov transformation can also be 
written as follows:
\begin{eqnarray*}
&&\frac{d {\mu}^1}{d{\mu}^2}=
\exp\Bigl\{
\int_{t_a}^{t_b}\Bigl[-\frac18
{\mu}^2\kappa\,
G^{RB}N^L_B N^M_R 
({\mathscr A}^{\alpha}_P\,K^P_{\alpha L}) \,
({\mathscr A}^{\beta}_D\,K^D_{\beta M})
\,\Bigr]dt+
\nonumber\\
&&\;\;\;\;\;\;\;\;\;\;\;\;\;\;\;\;\;\;\;\;\;\;\;+
\frac12{\mu}\,{\kappa}^{1/2}\,
N^L_K\,
({\mathscr A}^{\nu}_C\,K^C_{\nu L})
{\cal X}^K_{\bar M}\,dw_t^{\bar M}\,\Bigr]\Bigr\}\,.
\end{eqnarray*}
We see that the Jacobian can be presented  in terms of the projection operators and the "Coulomb connection" 
${\mathscr A}^{\nu}_P$. 

Thus, for the kernel of the evolution semigroup, we get
\begin{eqnarray*}
&&G_{\Sigma}(A^*_b,t_b;A^*_a,t_a)=\int_{ 
{\tilde{\xi}_{\Sigma}(t_a)=A^*_a}\atop
{\tilde{\xi}_{\Sigma}(t_b)=A^*_b}}
d\mu ^{\tilde{\xi}_{\Sigma}}\exp 
\left\{\frac 1{\mu
^2\kappa }\int_{t_a}^{t_b}
V(\tilde{\xi}_{\Sigma}(u))du\right\}
\nonumber\\
&&\exp\Bigl\{
\int_{t_a}^{t_b}\Bigl[-\frac18
{\mu}^2\kappa\,
G^{RB}N^L_B N^M_R 
({\mathscr A}^{\alpha}_P\,K^P_{\alpha L}) \,
({\mathscr A}^{\beta}_D\,K^D_{\beta M})\,\Bigr]dt+ 
    \nonumber\\
&&+\frac12{\mu}\,{\kappa}^{1/2}\,
N^L_K\,\left.
({\mathscr A}^{\nu}_C\,K^C_{\nu L})\,
{\cal X}^K_{\bar M}\,dw_t^{\bar M}\,\Bigr]\,
\right\},
\nonumber\\
&&(A^*=\pi_{\Sigma} (p))\,.
\end{eqnarray*}

\section{Conclusion}

Using the path integral measure factorization method, 
we have considered the  separation of the physical and unphisical
degrees of freedom in a pure Yang--Mills theory. 

Our path integrals have been formally defined 
by taking  the limits in the expressions obtained by  the  cylindrical approximations of  the local semigroups  that have been used 
by Belopolskaya and Daletskii  \cite{Dalecky} in their definition of  the path integrals given on a Hilbert manifold.
This definition is based on local stochastic processes 
given on charts of a manifold. In our case we have used  the (weak) Riemannian metric on the Hilbert manifold. Therefore  we were able   to deal only with the cylindrical versions of the true local stochastic processes which have their values in the Hilbert space of distribution. 

The factorization of the path integral measure has been  performed 
with the help of the nonlinear filtering stochastic
differential equation from the stochastic process theory.
As a result of our transformation we have obtained 
an integral relation (\ref{38}) between the fundamental solutions 
(the Green functions)  of the backward Kolmogorov equations that are given on the original manifold and on the manifold $\Sigma $ defined by the Coulomb gauge ${\partial}^k A^{\ast}_k({\mathbf x})=0$.
Both the left-hand side and the integrand of the right-hand side of this integral relation have been presented  in terms of the corresponding path integrals.

Considering the reduction onto the zero--momentum level 
(the $\lambda=0$ case), we have obtained the path integral 
representation of the Green function $G_{\Sigma}$. 
 The integrand of the path integral representing $G_{\Sigma}$ besides the potential term consists of the transformation Jacobian which has the functional dependence on the Coulomb connection.

The Green function $G_{\Sigma}$ is an Euclidean analog of  the Feynman propagator which is used to describe the quantum evolution 
on the orbit space. More precisely, it gives us an implicite description (in terms of  dependent gauge fields) of the gauge--invariant modes that could be associated with the  
``mysterious'' glueball particles and their excitations when
$\lambda \neq 0$.

In order to obtain these results, we have made  a number of  necessary assumptions, the main of which was about the 
possibility to expand the function, which depends on a ``group parameter'',
(the functional given on a gauge group), in a series 
over matrix elements of a special  irreducible representation 
of the gauge group used by Rossi and Testa. 
That is, we have supposed that there exists  an analog of  the  Peter--Weyl theorem for this representation.

We note that our assumptions appear to be justified, since they 
should lead  to the Schwinger quantum Hamiltonian 
in the physical subspace of the pure Yang--Mills fields.
But the final conclusion may only be done as a result of the examination of these questions. 

 Path integral transformations in our paper have been performed in 
 integrals defined over the formal measures.
In this connection, the basic question, which remains 
to be answered in further investigations,  is an account of the
regularization in our transformations (in case of using the regularized metric on a Hilbert manifold) and 
its possible influence on the final  structure of the
reduced Hamiltonian. 

\section{Acknouledgments}
The author would like to thank A. V. Razumov  for many useful discussions and V. E. Rochev for valuable advices. 

\section{Appendix}
\section*{Projection operators and their properties}

The projection operator onto the gauge surface $\Sigma $ 
can be presented in a following symbolical form: 
\[
(P_{\bot})^A_B={\delta}^A_B-{\chi}^{\mu}_B\;({\chi}
{\chi}^{\top})^{-1}{}^{\nu}_{\mu}\;({\chi}^{\top})^A_{\nu}\,.
\]
Let us consider an explicit form of this operator.

The matrix ${\chi}^{\mu}_B$ is given by the following formula:
\[
{\chi}^{(\nu, x)}_{\;\;\;\;(\alpha,i,y)}=
{\delta}^{\nu}_{\;\alpha}\,
\left[\,{\partial }_i(\mathbf x)\;{\delta}^3 (\mathbf x-\mathbf y)\,\right]\,.
\]
The transposed matrix $(\chi ^{\top})^{A}_{\mu}$, which is defined  by equality 
\[
(\chi ^{\top})^{A}_{\mu}=G^{AB}{\gamma}_
{\mu \nu}\chi ^{\nu}_{B},\,\,\, {\gamma}_{\mu \nu}
=K^{A}_{\mu}G_{AB}K^{B}_{\nu}\,,
\]
 has the following form:
\begin{eqnarray*}
(\chi ^{\top})^{(\alpha,m, x)}_{\;\;\;\;\;\;\;(\mu,z)}&=&
G^{(\alpha,m, x)(\beta,j,y)}{\gamma}_
{(\mu,z)(\alpha,u)}\chi ^{(\alpha,u)}_{\;\;\;\;(\beta,j,y)}
\nonumber\\
&=&\left[(\tilde{\cal D}\cdot {\cal
D})_{\mu}^{\alpha}(\mathbf z)\;{\partial}^m(\mathbf z)\;{\delta}^3
(\mathbf z-\mathbf x)\right]\,,
\end{eqnarray*}
where by $\tilde {\cal D}$ we denote the operator
\[
\bigl(-\delta ^{\varphi}_{\;\mu}{\partial}_k(\mathbf z)+c^{\varphi}_{\sigma
\mu}A^{\sigma}_k(\mathbf z)\bigr)\,.
\]
The ``product'' of the matrices $\bigl(\chi \cdot \chi ^{\top}\bigr)^
{(\nu,x)}_{\;\;\;\;(\mu,z)}$  
can be  written as follows:
\[
\chi ^{(\nu,x)}_{\;\;\;\;(\alpha, m,u)}
(\chi ^{\top})^{(\alpha,m, u)}_{\;\;\;\;\;\;(\mu,z)}=
\left[\bigl(\tilde{\cal D}_{\mu}\cdot {\cal D}^{\nu}\bigr)\,(\mathbf z)\,
\left(-{\partial}^m(\mathbf z)\,{\partial}_m(\mathbf z)\;
{\delta}^3(\mathbf z-\mathbf x)\;\right)\right].
\]
An inverse expression to this matrix product is 
\[
\bigl(\chi \cdot \chi ^{\top}\bigr)^{-1}
{}^{(\mu,z)}_{\;\;\;\;(\alpha,y)}=\int\; d^3u\;
\left[\bigl(\tilde{\cal D}\cdot
{\cal D}\bigr)^{-1}\right]^{\mu}_{\;\alpha}(\mathbf z-\mathbf u)\,\; K(\mathbf u-\mathbf y)\,,
\]
where $K(\mathbf x-\mathbf y)$ satisfies the following equation:
\[
(-1)\,{\partial}_m(\mathbf x)\,{\partial}^m(\mathbf x)\,K(\mathbf x-\mathbf y)={\delta}^3(\mathbf x-\mathbf y)\,.
\]
That is, we have
\[
\bigl(\chi \cdot \chi ^{\top}\bigr)
^{(\nu,x)}_{\;\;\;\;(\mu,z)}\;
\bigl(\chi \cdot \chi ^{\top}\bigr)^{-1}\;
{}^{(\mu,z)}_{\;\;\;\;(\alpha,y)}={\delta}^{\nu}_{\;\alpha}\;{\delta}^3(\mathbf x-
\mathbf y)\,,
\]
or
\[
\left(\tilde{\cal D}^{\nu}\cdot{\cal D}_{\mu}\right)(\mathbf z)\;
\left[\bigl(\tilde{\cal D}\cdot
{\cal D}\bigr)^{-1}\right]^{\mu}_{\;\alpha}(\mathbf z-\mathbf u)=
{\delta}^{\nu}_{\;\alpha}\;
{\delta}^3(\mathbf z-\mathbf u)\,.
\]
Thus  we  get that
\begin{eqnarray*}
&&\chi ^{(\alpha , x)}_{\;\;\;(\beta,m,z)}\;
\left[\bigl(\chi \cdot \chi ^{\top}\bigr)^{-1}\right]
^{(\epsilon ,u)}_{\;\;(\alpha ,x)}\;\bigl(\chi
^{\top}\bigr)^{(\mu,n,y)}_{\;\;\;\;(\epsilon, u)}=
\nonumber\\
&&{\delta}^{\mu}_{\,\beta}\;{\partial}_m(\mathbf z)\int d^3\tilde  y 
\,K(\tilde \mathbf y-\mathbf z)\;\left[{\partial}^n(\tilde \mathbf y)\;{\delta}^3 
(\tilde \mathbf y-\mathbf y)\right]\,.
\end{eqnarray*}
Taking this into account, we obtain
\begin{eqnarray*}
&&\bigl(P_{\bot}\bigr)^{(\alpha,k,x)}_{\;\;\;\;\;\;
(\beta,m,y)}=\nonumber\\
&&{\delta}^{\alpha}_{\,\beta}\left({\delta}^k_{\,m}\;
{\delta}^3(\mathbf y-\mathbf x)
+ {\partial}_m(\mathbf y)\int d^3 u\, K(\mathbf u-\mathbf y)\,\left({\partial}^k(\mathbf u)
\,{\delta}^3(\mathbf u-\mathbf x)\right)\;\right)
\end{eqnarray*}
Thus, the projection operator  can be written symbolically as
\[
\bigl(P_{\bot}\bigr)^{(\alpha,k,x)}_{\;\;\;\;\;\;(\beta,m,y)}=
{\delta}^{\alpha}_{\,\beta}\left[{\delta}^k_{\,m}+
{\partial}_m\frac{1}{(-{\partial}^2)}
{\partial}^k\right]
{\delta}^3(\mathbf y-\mathbf x)\,.
\]

In a finite dimensional case, the projector $N$ onto the subspace which is orthogonal to the 
Killing vector was defined  by 
the following formula:
\[
N^A_B={\delta}^A_B-K^A_{\alpha}(Q)\;
({\Phi}^{-1})^{\alpha}_{\mu}\;
{\chi}^{\mu}_B(Q)\,.
\]
In our case, it can be written as
\[
N^{(\alpha,i,x)}_{\;\;\;\;\;\;(\beta,j,y)}=
{\delta}^{(\alpha,i,x)}_{\;\;\;\;\;\;(\beta,j,y)}
-K^{(\alpha,i,x)}_{\;\;\;\;\;\;(\mu,z)}
\bigl({\Phi}^{-1}\bigr)^{(\mu,z)}_{\,\,\;\;(\nu,u)}
{\chi}^{(\nu,u)}_{\;\;\;\;(\beta,j,y)}\,.
\]
An explicit form of this projection operator is
\[
N^{(\alpha,i,x)}_{\;\;\;\;\;(\beta,j,y)}=
{\delta}^{\alpha}_{\;\beta}
{\delta}^{i}_{\;j}{\delta}^3(\mathbf x-\mathbf y)-
{\cal D}^{\alpha\,i}_{\epsilon}(\mathbf x)\int d^3 u
({\Phi}^{-1}){}^{(\epsilon,x)}_{\;\;\;\;(\beta,u)}(\mathbf x-\mathbf u)
\left[\partial _j(\mathbf u){\delta}^3 (\mathbf u-\mathbf y)\right]
\]
It  can also be written in a symbolical form:
\[
N^{(\alpha,i,x)}_{\;\;\;\;\;\;(\beta,j,y)}=
\left({\delta}^{\alpha}_{\;\beta}
{\delta}^{i}_{\;j}\,-{\cal D}^{\alpha\,i}\,
\left(\frac{1}{{\cal D}\,\partial}\right)_{\beta}\,
\partial _j\,\right)\,{\delta}^3(\mathbf x-\mathbf y)\,.
\]

The main properties of these projection operators are as follows:
\[
N^{(\alpha,i,x)}_{\;\;\;\;\;\;(\beta,j,y)}\;
N^{(\beta,j,y)}_{\;\;\;\;\;\;(\mu,k,z)}=
N^{(\alpha,i,x)}_{\;\;\;\;\;\;(\mu,k,z)}\,
\]
\[
N^{(\alpha,i,x)}_{\;\;\;\;\;\;(\beta,j,y)}\;
K^{(\beta,j,y)}_{\;\;\;\;(\mu,z)}=0\,
\]
\[
N^{(\alpha,i,x)}_{\;\;\;\;\;\;(\beta,j,y)}\;
\left(P_{\bot}\right)^{(\mu,k,z)}_{\;\;\;\;\;\;(\alpha,i,x)}=
N^{(\mu,k,z)}_{\;\;\;\;\;\;(\beta,j,y)}
\]
\[
\left(P_{\bot}\right)^{(\alpha,k,x)}_
{\;\;\;\;\;\;(\beta,m,y)}\;
N^{(\epsilon,n,z)}_{\;\;\;\;\;\;(\alpha,k,x)}=
\left(P_{\bot}\right)^{(\epsilon,n,z)}_
{\;\;\;\;\;\;(\beta,m,y)}
\]
\[
\left(P_{\bot}\right)^{(\alpha,k,x)}_
{\;\;\;\;\;\;(\beta,m,y)}\;\left(P_{\bot}\right)^
{(\beta,m,y)}_
{\;\;\;\;\;\;(\epsilon,n,z)}=
\left(P_{\bot}\right)^{(\alpha,k,x)}_
{\;\;\;\;\;\;(\epsilon,n,z)}
\]
\[
\left(P_{\bot}\right)^{(\alpha,k,x)}_
{\;\;\;\;\;\;(\beta,m,y)}\;
(\chi^{\top})^{(\beta,m,y)}_{\;\;\;\;(\mu,z)}=0
\]
\[
\left(P_{\bot}\right)^{(\alpha,k,x)}_
{\;\;\;\;\;\;(\beta,m,y)}\;
\chi^{(\nu,z)}_{\;\;(\alpha,k,x)}=0
\]
\[
N^{(\alpha,i,x)}_{\;\;\;\;\;\;(\beta,j,y)}\;
\chi^{(\nu,z)}_{\;\;(\alpha,i,x)}=0\,.
\]

\end{document}